\documentclass{emulateapj}
\usepackage{natbib}
\def\gsim{\;\rlap{\lower 2.5pt
 \hbox{$\sim$}}\raise 1.5pt\hbox{$>$}\;}
\def\lsim{\;\rlap{\lower 2.5pt
   \hbox{$\sim$}}\raise 1.5pt\hbox{$<$}\;}

\newcommand{\rms}{{\sc rms}}

\newcommand{\degree}{$^\circ$}

\newcommand{\Reff}{$R_{\rm eff}$}

\renewcommand{\arcsec}{$^{\prime\prime}$}
\renewcommand{\arcmin}{$^{\prime}$}
\newcommand{\kms}{\,{\rm km}\,{\rm s}$^{-1}$}

\newcommand{\othree}{[O~{\sc iii}]}

\newcommand{\Hone}{H{\small I}}

\newcommand{\PNS}{PN.S}

\newcommand{\Distance}{9.8}
\def\flux{erg~s$^{-1}$~cm$^{-2}$~\AA$^{-1}$}
\def\elec{$e^-$~s$^{-1}$~\AA$^{-1}$}

\slugcomment{revision 17 Feb 2007 - REVISION-1 }

\shorttitle{PN.S Data and Basic Dynamics for NGC~3379}
\shortauthors{Douglas et al.}

\begin{document}

\title{The PN.S Elliptical Galaxy Survey: Data Reduction, Planetary
  Nebula Catalog, and Basic Dynamics for NGC~3379 \footnote{Based on 
                       observations made with the William Herschel Telescope
                        operated on the island of La Palma by the 
                       Isaac Newton Group in the Spanish 
                        Observatorio del Roque de los Muchachos 
                        of the Instituto de Astrofisica de Canarias}}

\author{
N.G.~Douglas\altaffilmark{1}, 
N.R.~Napolitano\altaffilmark{2},
A.J.~Romanowsky\altaffilmark{3,1,4},
L.~Coccato\altaffilmark{1},
K.~Kuijken\altaffilmark{5,1},
M.R.~Merrifield\altaffilmark{4},
M.~Arnaboldi\altaffilmark{6,7},
O.~Gerhard\altaffilmark{8},
K.C.~Freeman\altaffilmark{9},
H.R.~Merrett\altaffilmark{4},
E.~Noordermeer\altaffilmark{4},
and M.~Capaccioli\altaffilmark{2}
}
\altaffiltext{1}{Kapteyn Astronomical Institute, University
of Groningen, the Netherlands}
\altaffiltext{2}{INAF-Osservatorio Astronomico di Capodimonte, 
Salita Moiariello 16,
80131, Naples, Italy}
\altaffiltext{3}{Departamento de F\'isica, Universidad de Concepci\'on, Casilla 160-C, Concepci\'on, Chile}
\altaffiltext{4}{School of Physics \& Astronomy, University of Nottingham, UK}
\altaffiltext{5}{University of Leiden, The Netherlands}
\altaffiltext{6}{ESO, Karl-Schwarzschild-Str. 2, D-85748 Garching Germany}
\altaffiltext{7}{INAF, Observatory of Turin, Strada Osservatorio 20, 10025 Pino Torinese, Italy}
\altaffiltext{8}{Max-Planck-Institut f\"ur Extraterrestrische Physik, Giessenbachstrasse,
D-85748 Garching Germany}
\altaffiltext{9}{Research School of Astronomy \& Astrophysics, ANU, Canberra, Australia}


\begin{abstract}
We present  results from Planetary Nebula Spectrograph
(PN.S) observations  of the elliptical galaxy NGC~3379
and a description of the data reduction pipeline.
We detected 214 planetary nebulae
of which 191 are ascribed to NGC~3379, and 23 to the
companion galaxy NGC~3384.  Comparison
with data from the literature 
show that the PN.S velocities have
an internal error of $\lesssim 20$\kms and a possible offset
of similar magnitude.  
We present the results of kinematic modelling and show that the
PN kinematics are consistent with
absorption-line data 
in the region where they  overlap.
The resulting combined kinematic data set, running from the center of
NGC~3379 out to more than seven effective radii (\Reff), reveals a mean
rotation velocity that is small compared to the random velocities, and
a dispersion profile that declines rapidly with radius.  
From a series of Jeans dynamical models we find the $B$-band mass-to-light
ratio inside five \Reff\ to be 
8 to 12 in solar  units, 
and  the dark matter fraction inside this radius 
to be less than 40\%.  
We compare  these and other results of dynamical
analysis with those of dark-matter-dominated merger simulations,
finding that significant discrepancies remain, 
reiterating the question of whether NGC~3379 has
the kind of dark matter halo that the current $\Lambda$CDM paradigm requires.
\end{abstract}

\section{Introduction}\label{sec:intro}
      The confirmation in the 1970s that dark matter
      dominates the mass of the Universe came about from dynamical studies
      of spiral galaxies \citep{1970ApJ...160..811F,1973ApJ...186..467O,
   	1978ApJ...225L.107R, 1986RSPTA.320..447V}.  In these systems, the 
	\Hone{} gas disks offer the
      	ideal diagnostic, since their extended nature allows one to probe 
      	very large radii where dark matter begins to dominate, and their cold
        disk-like structure ensures that the material
        is following approximately circular orbits, removing a major ambiguity
         in the study of its dynamics.
 
 A similar study of elliptical galaxies would be invaluable,
 as it would address such basic questions as
 whether the dark matter halos around these systems are similar to
 those around spirals, suggesting that the difference in observed
 morphology is just a matter of frippery, or that there are more
 fundamental differences between them. Unfortunately the paucity
of neutral hydrogen gas in ellipticals makes such a study observationally difficult.

Various dynamical tracers have been utilised to circumvent this problem.
For example, some ellipticals do contain
 significant amounts of gas, either in the form of rings of cold
material 
 \citep[e.g.][]{1993ApJ...416L..45B,1994ApJ...436..642F,2002AJ....123..729O}, 
 or as a fuzzy halo of hot 
 material \citep[e.g.][]{1994ApJ...427...86B}.
Neither is entirely satisfactory as a tracer, though, as these systems
are not usually what might be thought of as ``normal'' ellipticals:
 something peculiar must have happened to the galaxy to bestow it with
a gas ring, while X-ray emitting halos tend to only be measured in the
brightest systems, and are often confused with a surrounding
intra-cluster medium.  Other dynamical tracers such as globular
 clusters have also been employed \citep[e.g.][]{2006MNRAS.366.1253P},
 but the relatively modest numbers that have been observed to date in
 normal ellipticals stand no chance of solving for the full
 distribution function of these objects, in order to allow for all the
 possibilities in their orbital distribution.
Integrated-light
 absorption line spectroscopy has shown that the circular velocity profiles
 of ellipticals are typically flat to 2\Reff, and that in a
 fraction of the galaxies analysed the dark matter contributes $\sim
 10-40\%$ of the mass within \Reff\  when the stellar
 distribution has maximum mass, while in others no evidence for any
non-stellar matter is seen to 2\Reff\
 \citep[e.g.][]{2000A&AS..144...53K, 2001AJ....121.1936G, 
 2001MNRAS.322..702M, 2006ThomasPhD}.

For many years now, planetary nebulae (PNe) have been used to 
 extend stellar kinematic studies to larger radii, 
where the sought-after dark matter is 
expected to dominate the potential, and where the stellar orbital timescale
becomes long enough that some record of a galaxy's formation is likely
to have been preserved. Their bright  5007\AA\ \othree\ line is
relatively easy to detect and provides a line-of-sight velocity. 
Such measurements allow the velocities of stars to be measured
 out to arbitrarily large radii in elliptical galaxies; in fact, PNe
 become easier to detect at larger radii where the background light from
 the galaxy is fainter, so they neatly complement absorption-line
 studies at smaller radii.

 A number of attempts have been made to exploit this possibility,
 generally by using narrow-band imaging to identify PNe, and follow-up
 spectroscopy to obtain their velocities.  This approach has
been fruitful \citep[e.g.][]{1993ApJ...414..454C,1998ApJ...507..759A}, 
 but the two-step process, requiring precise astrometry, 
has tended to produce low yields.  

 \citet{1995ApJ...443L...5T} and \citet{2006AJ....131.2089S} 
have shown that one can detect
PNe and measure their velocities using a Fabry--Perot interferometer, 
now reaching sizable samples.  
Meanwhile, \citet{1999MNRAS.307..190D} developed the technique, coined 
``counter-dispersed imaging" ,
  in which images are obtained using a slitless 
spectrograph with the dispersive element rotated
by 180 degrees between exposures. 
 The emission lines
 from PNe appear as unresolved dots in each of the dispersed images, shifted in
 opposite directions by an amount proportional to their velocities.  By
 matching up the pairs of dots and measuring the distance between them, 
one can simultaneously identify PNe and measure their
 velocities.
 This technique was successfully implemented by \citet{2000MNRAS.316..795D}
 to study the kinematics of PNe in M94. 
 \citet{2001ApJ...563..135M} and
 \citet{2005ApJ...635..290T} subsequently used a dispersed-undispersed imaging
technique to measure the velocities of 531 PNe in NGC 4697 and 197 PNe
 in NGC 1344.
The Planetary Nebula Spectrograph again  utilizes counter-dispersed imaging   but
now two spectrographic cameras, fed by a beam-splitter arrangement at the 
grating module, are simultaneously in operation. Matched image
pairs, from the same collimated beam and via the same filter at the same
temperature, are therefore guaranteed. A more detailed description of the instrument
 is given in   \citet{2002PASP..114.1234D}, and Figure 11 of that paper shows
an example of the images obtained.

The \PNS\ has proved an effective tool for studying PN
kinematics in a variety of environments.  For example, we have
 completed a kinematic survey of M31, the resulting velocities of 2800
 PNe allowing us to model the dynamics of this system in unprecedented
 detail \citep{2006MNRAS.369..120M}.  However, the
 primary purpose of the instrument is to study the dynamics of a sample
 of ordinary early-type galaxies.  We commenced this core project during
commissioning of the instrument in 2001, with the intention of
 carrying out an extensive and definitive study of the large-scale
 dynamics of elliptical galaxies with a range of internal properties
 and in a variety of environments.  Our initial finding was unexpected.
The velocity dispersions of the first 
 three galaxies we had observed did not remain roughly constant with
 radius as the simplest dark halo models would predict; instead they
 went into a Keplerian decline that was consistent with these
systems having no dark halos at all \citep[hereafter R+03]{2003Sci...301.1696R}.  
Although such behaviour had been
 seen before \citep{1993ApJ...414..454C,2001ApJ...563..135M}, 
our findings provided the first strong indication that it might be a generic
 property of some types of ellipticals.

Subsequently, attention was drawn by \citet{2005Natur.437..707D}   
[hereafter D+05] to the well-documented degeneracy in the
study of velocity dispersion profiles, 
that one cannot unambiguously disentangle the effects of mass 
distribution and orbital
structure \citep{1982MNRAS.200..361B}: a declining profile could
 indicate a radial bias in the distribution of orbits instead of the
 lack of a dark halo.  (R+03) had
dealt with this possibility in some detail, and 
full orbit modeling  for the case of NGC~3379 
showed that, although some dark matter could be happily accommodated by this
ambiguity, the results were still inconsistent with the standard
cosmological predictions.  However, further analysis has
 indicated that $\Lambda$CDM (cold dark matter) halos are not entirely
 ruled out by the published data \citep[hereafter
 M{\L}05]{2005MNRAS.363..705M}, while studies of the kinematics of
 globular clusters around NGC~3379 seem to favor a more massive halo
 \citep{2006MNRAS.366.1253P,2006A&A...448..155B}.  D+05, citing the
 work of \citet{2004A&A...423..995M}, also introduced the worrying
 possibility that the detected, brighter, PNe might not be drawn from the
 stellar population as a whole, instead being skewed toward a
 non-representative younger sample of stars.  However, the observed
 universality of the bright end of the PN luminosity function 
\citep{1989ApJ...339...53C,2002ApJ...577...31C}
 and the
 lack of variation in kinematics with PN luminosity down to very faint
 limits in the disk of M31 would tend to argue that this is not the
 case \citep{2006MNRAS.369..120M}. 

 In R+03 space did not allow us 
 to present the full data reduction procedure, and so convince the
 reader that the novel PN.S instrument could be relied on to produce
 consistent and credible results, and that the inferred peculiar
 dynamics could not be attributed to some strange systematic problem
 with the unusual data analysis required.  This issue is particularly
 pertinent because at the time the paper was written we were still
 investigating the optimum procedure for processing PN.S data.  The
 full preliminary catalog of PN positions and velocities was also not
 published in the Science paper, making it difficult for others to
 reanalyze the dynamics using all the relevant information.

 It is the purpose  of this paper to address these issues. 
We will
 describe the now-finalized pipeline for analyzing PN.S data in some
 detail in \S\ref{sec:pipe}.  
We present the complete set of PN.S observational data for NGC~3379 in
 \S\ref{sec:data}.  Since a number of smaller PN data sets
 already exist for this galaxy, we have been able 
to calibrate the reliability of the PN.S pipeline,
as set out in \S\ref{sec:compare}.  
As a result of the improved data reduction
and a larger amount of integration time, 
the catalog for NGC~3379 presented here contains approximately
 double the number of PNe that went into the analysis of R+03.  We have
 therefore been able to make a more detailed analysis of issues such as
 completeness and contamination by other sources in
 \S\ref{sec:compcont}, as well as comparing the properties of the PNe
 to the more general stellar population in \S\ref{sec:starvPN}, to
 address the contentious question of how well PNe trace the bulk
stellar component.  Although this paper is not intended to provide a
 definitive study of the dynamics of NGC~3379, \S\ref{sec:vfield}
 presents a descriptive overview of its observed kinematics, while
 \S\ref{sec:dynamics} provides a more quantitative Jeans analysis of
 the galaxy's dynamics to re-examine the contentious questions
 regarding the nature of its dark halo, if any.  \S\ref{sec:squabble}
 takes another look at the question of whether or not the kinematic
 data from NGC~3379 is consistent with it having a standard
 $\Lambda$CDM dark halo as advocated by D+05, and we summarize the work
 in \S\ref{sec:summary}.

\section{The Reduction Pipeline}\label{sec:pipe}
As described in detail in \citet{2002PASP..114.1234D}, the PN.S is
mounted as a visitor instrument on the William Herschel 4.2m telescope,
operated by the Isaac Newton Group on the Island of La
Palma.
It effectively
comprises a pair of slitless spectrographs, dispersed in opposite
directions, with a common field of view of approximately $11'\times
10'$.  The dispersion is 1.29 pxl/\AA\ when used with the standard
CCD (see below) and the plate scale is 3.67 pxl/\arcsec
(somewhat less in the dispersed direction).
A narrow-band filter restricts the spectral range, so stars are
dispersed into short ``star trails" while an 
object such as a PN, with a single emission line within the filter passband, 
appears as an unresolved dot in each of the two arms of the
spectrograph, 
which we have arbitrarily labeled ``left'' and
``right''.

In addition to conventional bias frames, sky flats, dome flats and 
flux calibrations with the aid of standard stars, 
we also obtain images through a mask that
we can insert into the beam. The mask contains an array of 178
0.9\arcsec{} holes at calibrated locations, and is illuminated
with either a tungsten continuum lamp to determine the filter bandpass
or a Cu-Ne-Ar lamp to determine the wavelength solution and spatial
distortions.

Data reduction is via a dedicated
pipeline written in the {\sc iraf}\footnote{{\sc iraf} is
distributed by NOAO, which is operated by AURA Inc., under contract
with the National Science Foundation.} script language, with some
additional routines written in {\sc fortran}.
Because of its
non-standard nature and its importance to the interpretation of the
derived PN kinematics, we now describe the pipeline processes in some
detail.  

\subsection{Initial Processing}
The \PNS\ uses as its detectors a pair of 
2k$\times$4k 13.5$\mu m$ pixel EEV CCDs maintained by the 
Isaac Newton Group itself, simplifying the integration of the PN.S
within the Observatory's data acquisition system.  The
first step in data reduction is 
to trim the data frames to
the $2154 \times 2500$ pixels that are illuminated by the
spectrograph.  There is little structure in the bias levels of these
CCDs, so debiassing is achieved by subtracting a polynomial surface
fitted to the under- and over-scan regions of the CCD.  
Bad pixels on the CCDs are identified using unsharp masking of sky or
dome flats. 
The flats are combined to enhance their signal-to-noise
ratio and remove cosmic rays, and are then compared to a
median-smoothed version of the averaged flat.  Any pixels that are
deviant by more than 9$\sigma$ are flagged as bad.  In addition, we
maintain a library of known defects in these chips.  From these bad
pixel lists, we construct a binary mask of defects, and interpolate
over bad columns using the {\sc iraf} task {\sc fixpix}.

\subsection{Cosmic Ray Removal}
We have experimented with different methods for the removal of cosmic
rays.  In the M31 survey \citep{2006MNRAS.369..120M}, the {\sc
l.a.cosmic} algorithm \citep{2001PASP..113.1420V} was used successfully.
However, for  longer integration times 
several iterations were required for even an imperfect
correction
which left  residuals 
that could be confused with objects.  
We therefore
devised a simpler routine, {\sc crcleankk}, which examines 
each pixel's neighbors to see if the
gradient to these adjacent pixels is consistent with the local
brightness and the seeing under which the data were obtained.  Any
pixels where this gradient is found to be too high are flagged as
cosmic rays and removed.

\subsection{Flatfielding}
Sensitivity variations from pixel to pixel are calibrated using
twilight sky flats.  A customized script, {\sc ppcorrect}, first
creates a series of normalised 
flat field images by dividing each sky exposure by
a median-filtered version of itself.  These flats are then combined
with a rejection threshold to eliminate residual cosmic rays, and
weighted to maintain Poisson error characteristics.  The resulting
master flats for the left and right arms of the spectrograph are then
applied to all the science and calibration frames to remove
pixel-to-pixel variations.

The overall vignetting of the spectrograph is a little harder to
quantify.  The PN.S uses a slow shutter that takes $\sim 15$
seconds to open or close, so the effective exposure time varies
somewhat across the field.  This effect is completely negligible for
science exposures, with typical integration times of a half an hour, 
but does affect the illumination of short
twilight sky exposures significantly, so these data cannot be
used to calibrate the vignetting function of the telescope. Instead,
we measure the over-all illumination pattern using 
dome flats, for which the slow shutter is not an issue
as the exposure time is determined by switching the lamps on and off.
 One problem with such flats is that the illumination of
the inside of the dome tends to contain gradients due to the placement
of the illuminating lamps.  We have overcome this issue by combining
dome flats obtained with the instrument rotator at a number of
position angles, thus averaging away any such gradients.

\subsection{Wavelength Calibration and Image Rectification}\label{sec:wcal}
We now reach the heart of the pipeline.  What most complicates the
analysis of these slitless spectroscopic data is that, at least in
one dimension, position on the CCD does not map directly from a single
location on the sky: the dispersive element means that this coordinate
encodes a combination of position and wavelength.  
To disentangle
these effects, we utilise the Cu-Ne-Ar arc lamps of the
calibration unit of the WHT, close to the focal plane, 
obtaining the images of several lines for each of the
178 apertures in the calibration mask. The instrumental
setup is as close as possible to that during PNe observations,
and includes the same narrow-band filter. This limits the number of
arc lines seen to $\sim 5$, which is however sufficient.  
An automated task, {\sc alignspots}, 
locates and centroids the spots on the CCD, and feeds these data to
a routine which fits a joint polynomial fit for position and wavelength.
The fitting orders used have  been standardized by experiment, and residuals
can be inspected graphically, drawing attention to systematic errors. 
The solution represents a map from particular positions in the sky
reference frame, $\{X,Y\}$, and wavelength, $\lambda$, to a position
on the CCD, $\{x,y\}$:
\begin{equation}\label{eq:map}
\{x,y\} = f(X, Y, \lambda).
\end{equation}

This mapping is, of course, not one-to-one, since different
combinations of position and wavelength will fall in the same CCD
pixel, but the essence of the counter-dispersed technique is that the
two arms of the spectrograph provide two independent mappings to
positions on the two CCDs, $\{x_L,y_L\}$ and $\{x_R,y_R\}$, both of
which are calibrated as above.  As long as any pairs of detected
sources can be matched up unambiguously, the mappings that we have
obtained from equation~\ref{eq:map} can be uniquely inverted to transform
\begin{equation}\label{eq:unmap}
\{x_L, y_L, x_R, y_R\} \rightarrow \{X,Y,\lambda\}.
\end{equation}
This inversion is 
ultimately performed by a pipeline script called
{\sc pnsvel}, which also converts the wavelength of the
emission-line source into a velocity, under the assumption that the
line in question is the \othree\ one with a rest wavelength of
5006.8\AA.  These velocities are 
converted to heliocentric values
using the {\sc iraf} task {\sc rvcorrect}.
The mapping functions of equation~\ref{eq:unmap} also allow us to
rectify the spatial distortions introduced by the spectrograph, and
transform the images to ``true'' sky coordinates, using the {\sc iraf}
commands {\sc geomap} and {\sc geotrans}.  For any individual
image, this transformation is not really necessary since we have all
the information for recovering sky coordinates encoded in
equation~\ref{eq:unmap}, but it is required for the next step of
combining exposures taken at different times.

\subsection{Image Combining}\label{sec:pipecomb}
Once the individual images from the two arms are rectified, they can
be aligned and added to produce the final left and right image pair.
It is possible that there might also be some degree of rotation or
change in image scale between exposures obtained over a series of
nights,  
or even from different observing seasons; we have allowed for
this possibility in our routine to combine images, but have found in
practice that it does not occur at a significant level.

One challenge here is that we do not have any bright point sources
that we can use to register the different images.  Even the brightest
PNe are barely detectable in individual frames, so cannot be used to
determine the offsets.  
Instead, we 
utilise the  star trails described earlier
in this section to align images.
The locations of stars are determined using a
task called {\sc xstartrails}, which cross-correlates the star
trails in the images with simulated trails, to 
determine their centroids.  Given the greater extent of the star trails in the
spectral direction, we might expect the coordinates to be less well
determined in this direction, but since the offsets are determined
using many 
(typically 10-15) star trails, 
positional accuracy is not significantly limited by this effect (see
\S\ref{sec:compare}).

One slight subtlety in this alignment process is that the sub-pixel
shifting of images to a common reference frame requires an
interpolation of the data.  Some information is lost with each such
interpolation, and the image rectification  applied in
{\S\ref{sec:wcal}} means that we have already made one such
transformation.  Rather than applying a further transformation, we
instead go back to the original unrectified images, and apply a single
transformation that encompasses both the geometric correction and the
shift to a common reference frame.

In addition to finding relative offsets between images, {\sc
xstartrails} can also be used to determine an absolute astrometric
reference frame for the data.  It does so by reading in the positions
of a number of suitable stars for the appropriate field from the
USNO-B catalog \citep{2003AJ....125..984M} and applying the proper
motion correction to the epoch of the observation.  In a typical
high-latitude field, there are 30 -- 40 stars in this catalog at
magnitudes $B < 20$, which cause bright enough star trails to be
easily detected.  The approach to wavelength calibration that we have
adopted means that the center of the star trail is mapped onto the CCD
via equation~\ref{eq:map} at a location corresponding to the star's
sky coordinates and a wavelength given by the central wavelength of
the narrow-band filter.  We can therefore use the observed centroid of
the star trail to define the absolute astrometric frame, although the
poorer definition of the star trail's centroid in the dispersed
direction means that we would once again expect the reference frame to
be less well determined in this direction.  Fitting the observed
locations to those in the USNO-B catalog with a low-order polynomial
plate solution (quadratic in the dispersed direction and linear in the
undispersed direction) gives a satisfactory fit with typical residuals
of 1 -- 2 arcseconds.  Since the solution uses 30 -- 40 stars, the
random error from the fit will be significantly smaller.  As we will
see in \S\ref{sec:compare}, absolute coordinates are tied down by this
procedure to better than an arcsecond in both the dispersed and
undispersed directions, which is entirely adequate for this survey.

In order to combine the individual data frames appropriately, {\sc
xstartrails} also evaluates the quality of each exposure to determine
how much weight it should have in the final co-added image.
Specifically, the task estimates the seeing full-width at half
maximum, $\varsigma$, by taking a cut through the star trails in the
undispersed direction, it determines the relative transparency,
$\tau$, by comparing the brightnesses of star trails in different
images (allowing for differences in exposure time, $t$), and it
measures the sky background level, $B$, around the trails.  To account
for these quality indicators in creating a combined image, each frame
is weighted by a factor
\begin{equation}
W = {t \tau \over \varsigma^2 B}. 
\end{equation} 
With this scaling, images with the same quality but different exposure
times will be weighted equally, since $B \propto t$, and the final
combined images for each arm optimally combine data taken in
variable conditions.  

\subsection{Photometry}\label{sec:pipephot}
Although not primarily designed as a photometric instrument, the PN.S
does provide some information on the brightnesses of the objects that
it detects.  Since the PN luminosity function is broad, and we
typically detect PNe over a range of several magnitudes, even quite
crude photometric measures are useful in, for example, trying to see
if their kinematics vary with luminosity.
To determine the absolute zero-point of the photometry, we observe a
number of flux standard stars throughout each observing run.  Because
of the slow shutter speed, we adopted a different observational set-up
for  standards: leaving the shutter open, we let the time between
the clear and the read-out of the CCD determine the integration time.  The
CCDs are windowed for this procedure, so there is no significant contamination
of the standard star from additional flux during read-out.

As with any other star, the light from the standard star is dispersed
by the PN.S into a short filter-limited star trail. 
The counts in each such trail are integrated 
over a small number of pixels in the
dispersed direction at a point in the center of the trail
corresponding to the peak of the filter transmission.  These values
are then turned into a photon detection rate by subtracting background
counts derived from 
pixels surrounding the trail, converting from ADUs to
electrons as appropriate for the CCD, correcting for the airmass of
the observation, and dividing by the exposure time to get a rate.
This count rate can then be compared to the fluxes provided by
\citet{1990AJ.....99.1621O} and \citet{1982ApJS...48..395S} to find
the photometric conversion factor.  Since the optical paths to the two
arms differ, this process is carried out separately for each arm of
the spectrograph.

One minor complication in applying this calibration to detected
emission-line 
sources is that the observed counts will depend on the wavelength at
which the source is detected: the more a source is Doppler-shifted 
towards an edge of the filter's narrow bandpass, the more light will be lost.
Further, the bandpass of the filter changes somewhat depending on the
angle of the light's incidence; since the filter is in the collimated
beam, the bandpass will vary with position in the field of view.  We
therefore, during daytime, obtain long-exposure continuum lamp 
images through the mask
in order to map out these effects.  This calibration
procedure  is a significant
improvement over what is possible in traditional narrow-band surveys,
where the same effects occur, but
where detailed determination of the bandpass is more problematic.  In
practice, the profile of the filter and the narrow range of
angles involved mean that the correction that we apply to source
fluxes (once their positions and wavelengths have been determined) is
never more than $\sim 15\%$.

The other complicating factor is that observations can be undertaken at
a wide range of air masses, and under varying photometric conditions. 
Fortunately, it is fairly straightforward to account for this
by comparing the measured counts in the star trails with the
counts obtained during the best conditions. Once again this task
is carried out automatically using {\sc xstartrails}. 

Once these flux calibrations have been applied to the detected
emission-line objects, the derived values can be converted to the
usual $V$-band equivalent magnitude.  To do so, it is assumed that the
flux, $F_{\rm 5007}$, originates from the \othree{} line at 5007\AA{}, in which
case the equivalent magnitude is given by 
\begin{equation}\label{eq:m5007}
m_{5007} \equiv 
    -2.5 \log[F_{\rm 5007}({\rm erg}\,{\rm s}^{-1}\,{\rm cm}^{-2})] - 13.74
\end{equation}
\citep{1989ApJ...339...39J}.

\subsection{Object Finding}\label{sec:find}
Perhaps the single most important element in the pipeline is the
detection of sources in the final images.  A careful balance must be
struck between being too optimistic and identifying noise spikes as
objects, and being too pessimistic and throwing out valid sources from
the ultimate catalog.  An extra factor here is that a source must be
detected in both arms of the spectrograph to return a velocity, so we
need also to consider the optimum way to process the data to obtain
coincident detections.  We have therefore invested considerable
effort into optimizing this process, looking at both visual inspection
and semi-automated techniques.  Our conclusion is that both approaches
have their merits: the automated algorithms are better at assessing
the significance of faint object ``detections,'' while the human
approach is better at dealing with non-standard situations such as an
object very close to a star trail.  We have therefore opted to perform
both types of analysis, each separately carried out by two
researchers, resulting in a total of four independently-derived
catalogs.  Once the separate lists are produced in this independent
manner, they can be compared to assess their efficiencies, and any
discrepancies can be dealt with via an agreed dispute resolution
process.

For the automated procedure, initial identification of sources is
carried out using SExtractor \citep{1996A&AS..117..393B}.  To reduce
the effects of large-scale diffuse light from the galaxy, the stacked
image obtained from each arm is first median subtracted, using a $49
\times 49$ pixel kernel to define the median around each pixel.  The
input parameters for SExtractor were tested on artificial images in
order to find an optimal compromise between object finding and
spurious detections. 
Groups of at least 5 contiguous pixels whose flux
above the background is at least 0.7 times the \rms\ of the background are
classified as potential sources. 
The choice of criteria for separating point-like
PNe from extended background sources is left to the individual
researcher carrying out the search, as different sets of criteria seem
to produce similar results, and we want to test how sensitive the
results are to the specific criteria adopted.
Spurious objects are eliminated
by  cross correlation of the catalogues produced
for the left and right arm respectively.
The researchers using
the automated procedure then inspect the identified sources by eye to
make sure that the algorithm has not been fooled by star trails or
other spurious features.

The researchers who draw the short straws carry out the entire
procedure by eye, laboriously going over every part of the images from
the two arms of the spectrograph, blinking between them to look for
pairs of point-like objects that appear at the appropriate locations
to be the signature of a PN.

Once each catalog of potential emission-line objects has been compiled
for the left and right arms of the instrument, they can then be
matched up in an automated algorithm to convert their locations on the
two CCDs into a position on the sky and a velocity, as in
equation~\ref{eq:unmap}.  The primary criterion for searching is that
the detections in the two arms have to lie at the same coordinate in
the undispersed direction to within two pixels (set to accommodate the
maximum positional uncertainty in the faintest sources).  
 In the
dispersed direction, the coordinates must match up to within 200
pixels, since a separation greater than this would correspond to
a wavelength outside the filter bandpass.  
In the
small number of cases where the field is sufficiently crowded that
more than one match between pairs is possible, we use the fluxes of
the sources to resolve the ambiguity  where possible.  
In the end, there are very few apparent detections which are not
allocated to a unique matched pair.
From the resulting four independent lists of objects, a ``core
catalog'' is constructed where all four agree that a source had been
detected.  A ``dispute catalog'' contains the remaining sources that
were not found in all four searches.  These disputed objects are all
inspected independently by the four catalogers.  Those cases where all
now agree that the source is real are combined with the core catalog
to produce a ``consensus catalog.''  Objects that are 
significantly extended are placed in the ``extended
catalog'' for later study, and for exclusion from the PN analysis.
Finally, to produce the most uniform data set possible, a
signal-to-noise ratio cut of five (in each arm) is applied to produce
the ``final catalog.''  

\section{Application to NGC~3379 Observations}\label{sec:data}

As a first application of the data reduction pipeline, we have
analyzed PN.S observations of NGC~3379.  This galaxy provides a good
initial test case for a number of reasons.  As an E1 galaxy at a
distance of \Distance~Mpc \citep{2003ApJ...583..712J}, 
with an absolute magnitude of 
$M_B=-19.8$  and an effective radius of \Reff =47\arcsec\ (see \S \ref{sec:stellarmod})
corresponding
to 2.2 kpc, it is fairly typical of the systems in our survey of
ordinary elliptical galaxies.  
Also, since NGC~3379 was one of the objects in R+03, we can
check that the initial results are confirmed, and can see how much
better we can do with the fully-developed pipeline than was possible
with our earlier procedure.  There also exists a number of 
smaller 
PN data sets from NGC~3379, which we can compare to our results
as a test of the reliability of the PN.S and
its pipeline-processing.  As Table~\ref{tab:obs} shows, we have
obtained some 19 hours of integration on this object under a variety
of seeing conditions over multiple observing seasons, so we can also
test our ability to combine such heterogeneous data into an effective
stacked image.  Using the scaling relationship that point-source
detectability goes as the inverse square of the seeing, these
observations are the equivalent of 7 hours observing in nominal
one arcsecond seeing, which is typical of  the depth to which our elliptical galaxy
survey is being made.

\begin{table}                                         
\caption{Log of NGC~3379 Observations}\label{tab:obs}
\begin{center}
\setlength\tabcolsep{5pt}
\begin{tabular}{lll}
\hline \hline
\noalign{\smallskip}
Date         & Integration &  Seeing \\
             & (hours)     &  (FWHM) \\
\hline
2002 March 7 & 2.0         & 1.9\arcsec \\  
2002 March 8 & 4.8         & 2.5\arcsec \\
2002 March 9 & 1.5         & 1.4\arcsec  \\
2002 March 10& 2.5         & 1.4\arcsec \\
2002 March 14& 2.6         & 2.5\arcsec\\
2003 March 1 & 2.0         & 1.1\arcsec \\
2003 March 4 & 1.6         & 1.7\arcsec \\
2003 March 6 & 0.3         & 1.2\arcsec \\
2003 March 7 & 1.5         & 1.2\arcsec\\
\hline
\end{tabular}
\end{center}
\end{table}

NGC~3379 has a systemic velocity of 910\kms, so
the observations were made using our ``B'' narrow-band filter, which
has a default nominal central wavelength of $5033.9$~\AA{} and FWHM of
31~\AA{}.  This central wavelength was shifted to $5026.74$~\AA{} by
tilting the filter by 6 degrees, as described in
\citet{2002PASP..114.1234D}. One side effect of tilting the filter is
that the bandpass varies more significantly with position in the field
of view.  To compensate for this effect, the pointing was offset by
2\arcmin{} from the center of the field, but the relatively small size
of the galaxy and the 10\arcmin{} field of PN.S mean that the
observations still reach easily to beyond 7~\Reff. 
The
instrument position angle was set to $90^\circ$, such that the grating
dispersed in the East-West direction.

\begin{table}
\caption{Flux Calibration}\label{tab:fluxstd}
\begin{center}
\noindent{\smallskip}\\
\begin{tabular}{l l l l}
\hline
Star Name  &  $C_L$                  &  $C_R$                 &  $E_{L+R}$ \\
(1)        &  (2)                    &  (2)                   &  (3)       \\   
\hline 
Feige 67   &  $2.22 \times 10^{-16}$ & $2.04 \times 10^{-16}$ &    0.27 \\
Feige 67   &  $2.26 \times 10^{-16}$ & $2.07 \times 10^{-16}$ &    0.26 \\
Feige 98   &  $2.07 \times 10^{-16}$ & $1.92 \times 10^{-16}$ &    0.29 \\
Feige 98   &  $2.10 \times 10^{-16}$ & $1.91 \times 10^{-16}$ &    0.28 \\
\hline
Median     &  $2.16 \times 10^{-16}$ & $1.98 \times 10^{-16}$ &    0.28 \\
\hline 
\end{tabular}
\\
\begin{minipage}{8.5cm}

NOTES -- (1): Flux standard's name.
         (2): Conversion factors in units of \flux\ / \elec\ for the
              left and right arms. 
              The median value is shown at the bottom of the table.
         (3): Total efficiency of the instrument, 
              $E_{L+R}=\frac{\epsilon}{(C_L + C_R) S}$, where
              $\epsilon=3.95 \times 10^{-12}\,{\rm erg}$ is the energy per photon
               at the observed wavelength of 5026.74 \AA, and $S$ is the total 
              collecting area of the telescope.
\end{minipage}

\end{center}
\noindent{\smallskip}\\
\end{table}

We also observed a number of flux standard stars and carried out the
photometric calibration as set out in \S\ref{sec:pipephot}.  The
results are presented in Table~\ref{tab:fluxstd}, which also gives a
sense of the level of uncertainty in the photometric calibration of
\PNS\ data.  The table additionally shows the derived total efficiency
of telescope-plus-instrument, which lies close to the design
specification of 30\%.

\begin{table*}
\caption{Catalog of PNe in NGC~3379}
\begin{center}
\setlength\tabcolsep{5pt}
\begin{tabular}{lllc llcc}
\hline \hline
\noalign{\smallskip}
   ID         &   RA       &  Dec          &    wavelength     &    $v_{\rm helio}$   &   $m_{\rm 5007}$ &      SW06  &    CJD93  \\
 PNS-EPN-     &   J2000    &  J2000        &    \AA            &    \kms              &                  &            &        \\
\hline\\      
 NGC3379 1   &  10:47:31.22  & 12:39:20.3  &   5023.7   &      1005   &   27.04   &   -- 	 &    --  \\
 NGC3379 2   &  10:47:32.23  & 12:32:09.9  &   5022.1   &       909   &   26.55   &   -- 	 &    --  \\
 NGC3379 3   &  10:47:33.61  & 12:34:08.7  &   5021.8   &       892   &   27.67   &   -- 	 &    --  \\
 NGC3379 4   &  10:47:34.43  & 12:35:54.1  &   5022.9   &       958   &   26.89   &   -- 	 &    --  \\
 NGC3379 5   &  10:47:34.83  & 12:34:40.4  &   5024.3   &      1038   &   27.83   &   -- 	 &    --  \\
 NGC3379 6   &  10:47:35.27  & 12:36:48.1  &   5021.5   &       874   &   27.15   &   -- 	 &    --  \\
 NGC3379 7   &  10:47:35.96  & 12:33:51.1  &   5024.0   &      1024   &   26.95   &   -- 	 &    --  \\
 NGC3379 8   &  10:47:36.35  & 12:33:54.0  &   5024.0   &      1022   &   26.51   &   -- 	 &  77   \\
 NGC3379 9   &  10:47:36.44  & 12:33:05.8  &   5022.1   &       909   &   27.10   &   -- 	 &    --  \\
 NGC3379 10   &  10:47:36.59  & 12:33:50.3  &   5024.4   &      1043   &   24.99   &   -- 	 &    --  \\
 NGC3379 11   &  10:47:37.97  & 12:35:09.1  &   5021.8   &       890   &   27.99   &   -- 	 &    --  \\
 NGC3379 12   &  10:47:38.06  & 12:34:58.1  &   5023.1   &       967   &   26.14   &   -- 	 &  54   \\
\multicolumn{1}{c}{$\vdots$}    &
\multicolumn{1}{c}{$\vdots$}    &
\multicolumn{1}{c}{$\vdots$}    &
\multicolumn{1}{c}{$\vdots$}    &
\multicolumn{1}{c}{$\vdots$}    &
\multicolumn{1}{c}{$\vdots$}    &
\multicolumn{1}{c}{$\vdots$}    &
\multicolumn{1}{c}{$\vdots$}    \\
 NGC3379 212   &  10:48:09.93  & 12:37:33.4  &   5020.7   &       824   &    8.63   &   -- 	 &    --  \\
 NGC3379 213   &  10:48:10.08  & 12:35:49.8  &   5023.0   &       961   &   27.12   &   -- 	 &    --  \\
 NGC3379 214   &  10:48:10.58  & 12:37:23.4  &   5021.4   &       868   &   26.39   &   -- 	 &    --  \\
\hline\\
\end{tabular}\label{tab:cat}
\end{center}
\begin{minipage}{16 cm}
NOTES - The \PNS\ data follow the usual definitions, and the ID is
chosen to be compatible with proposed IAU standards (EPN standing for
``Extragalactic Planetary Nebula").  The absence of a magnitude entry
indicates that a reliable value could not be ascertained.
 Daggers indicate PNe which were allocated to NGC~3384 (see \S~\ref{sec:n3384}).
 Columns SW06 and CJD93 show matching PNe in the catalogs of
\citet{2006AJ....131.2089S}~and \citet{1993ApJ...414..454C}\
respectively.  As is clearly indicated, two of the matches with
\citet{2006AJ....131.2089S} are with PNe from their table 6 (NGC3384
catalogue).
 Objects were considered matched with a PN in our catalog if the
positions agreed within 3.0\arcsec\ and their velocities within 100\kms\
without adjustment for systematic offsets in either case.
 The full catalogue is available in the associated electronic material.
\end{minipage}
\end{table*}

The NGC~3379 observations were pipeline-processed as described in
\S\ref{sec:pipe}, resulting in a consensus
catalog of 214 objects, which have an overwhelming likelihood
of being PNe.
The dispute resolution 
procedure also produced a catalogue of 17 uncertain PNe.  Of
these, four had uncertain velocities because the pairs of objects
in the two arms of the spectrograph could not be matched up
unambiguously.  Although such objects cannot be used in the kinematic
studies, they are nonetheless firm 
detections, so must be included in
other analyses such as the number count distribution.  The final
signal-to-noise ratio cut did not exclude any of the consensus
detections, so the final product, the catalog of PNe, contains
214 PNe as presented in
Table~\ref{tab:cat}.

\section{Comparison with Existing Data}\label{sec:compare}
One of the reasons for basing this initial application on the
observations of NGC~3379 was that several smaller PN kinematic data
sets already exist for this galaxy, so we can obtain an independent
measure of the reliability of the data set.  In addition to our own
preliminary analysis of \PNS\ data, there are catalogs in the
literature from \citet{1989ApJ...344..715C}, \citet{1993ApJ...414..454C} and
\citet{2006AJ....131.2089S}, which we now consider in turn.

\subsection{Comparison with Romanowsky et al.\ (2003)}
A simple internal check on the robustness of results derived from \PNS\
data is provided by a comparison of the current analysis with the ad
hoc preliminary analysis that we performed on the same data in R+03.
That preliminary analysis picked out 109 PNe in a first pass through
the 2002 data, and the \PNS\ catalog matches 106 of these objects.
The more detailed analysis in this paper has flagged the remaining
three objects as extended, thus excluding them from the catalog.  
The new pipeline produces velocities with a positive offset of
34\kms\ from the earlier analysis,
probably because the wavelength calibration can now be performed
more accurately. The joint scatter for the 106 mutual objects  
is 22\kms.  
Neither of these velocities is large enough to be relevant to
the scientific results from the instrument.

\subsection{Comparison with Ciardullo et al.\ (1989,1993)}

\citet{1989ApJ...344..715C} measured positions and magnitudes for 93 PNe
in NGC~3379 using the traditional on-band/off-band imaging technique,
and then \citet{1993ApJ...414..454C} followed this work up with velocity
measurements of 29 of the PNe using multi-fiber spectroscopy.  Comparing
\PNS\ coordinates to these published values, we can with reasonable
confidence  match up 76 of the 93 sources on the basis of position (with
a 3\arcsec search window), utilising radial velocity data where
available to decide between ambiguous matches. If radial velocities are
not available we take the best match on position and magnitude.  Of the
17 PNe which were not matched by objects in our survey three were found
to have been rejected as extended, five were embedded in star trails or
lost in the galaxy's diffuse light, and the remaining nine apparently
lay below
 the magnitude limit of our data.
Of the subset of 29 objects for which velocities
were obtained by C+93, we are able to match 26 on the basis of position alone,
using radial velocity data to eliminate just one ambigous match 
(PNS-109 with C+93-64).
As expected, the differences in position  are marginally
larger in right ascension than they are in declination, as the stars from
which our astrometric solution was derived are dispersed in this
direction (see \S\ref{sec:pipecomb}).  However, the combined \rms\ difference
in position is only 1.6\arcsec, which is completely
negligible in studying the dynamics of a system of this size.
\begin{figure}
\includegraphics[width=8.6cm]{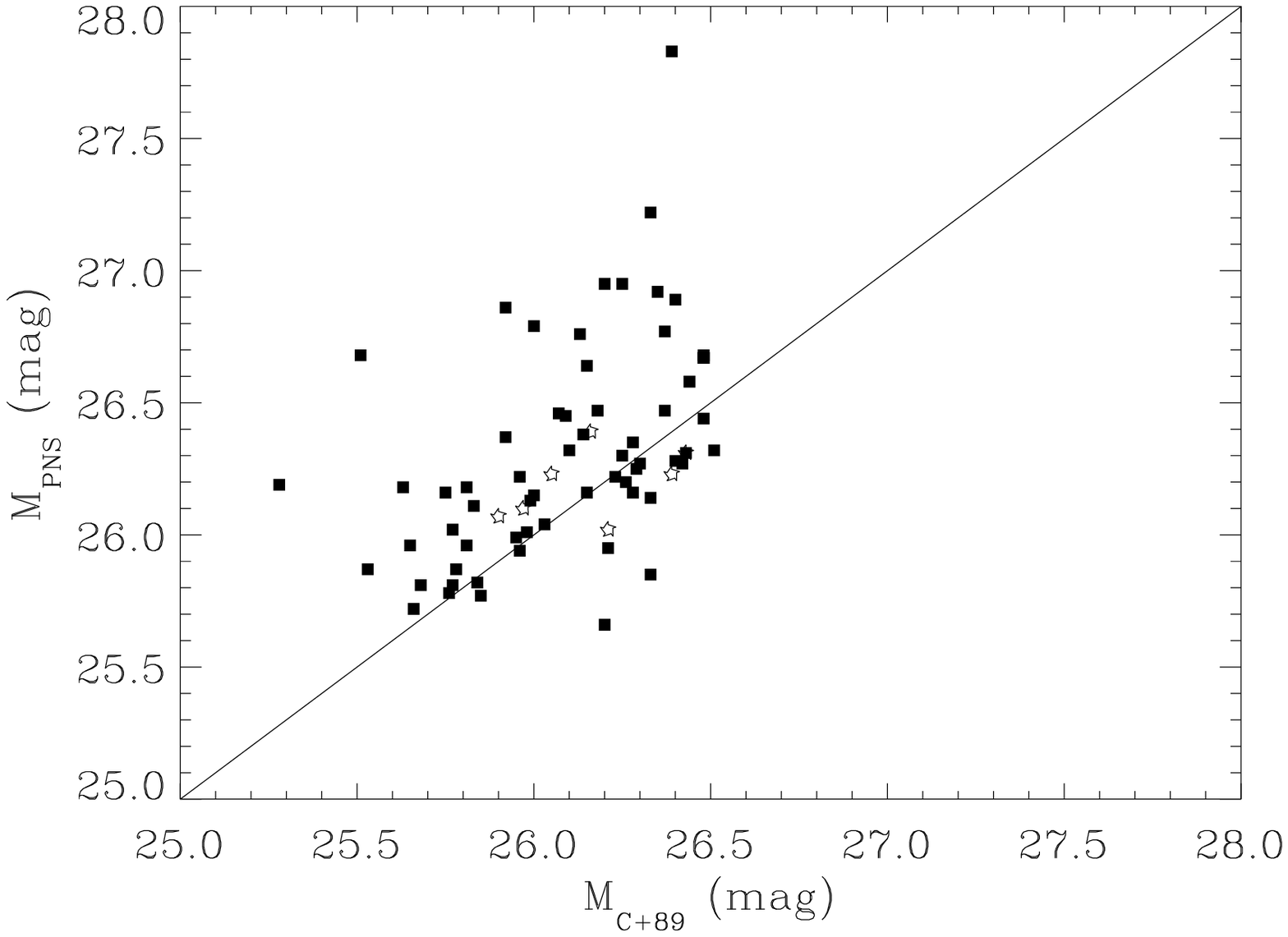}
\includegraphics[width=8.6cm]{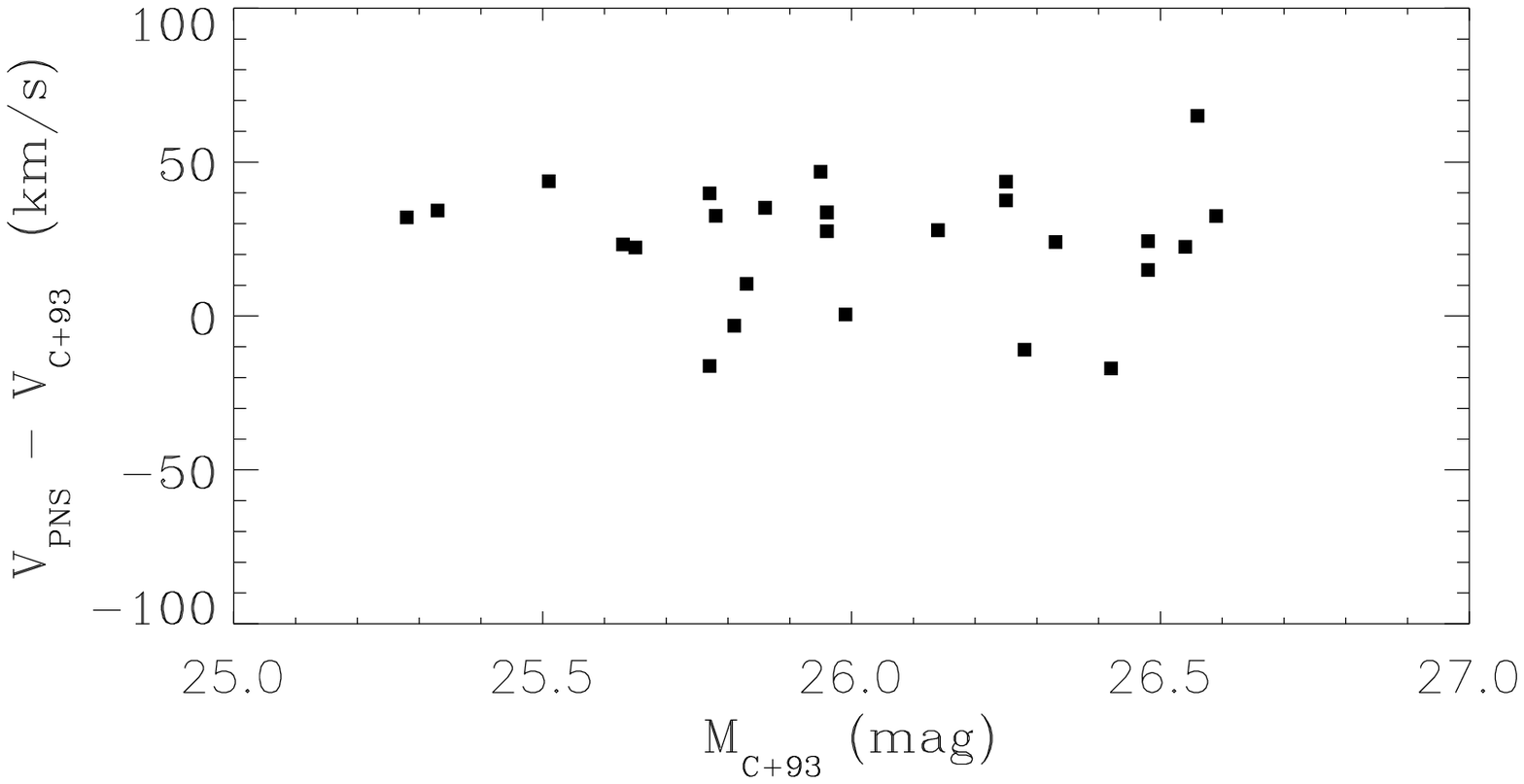}
\caption{Comparison between the measured magnitudes (upper panel)
and velocities (lower panel) for the PNe common to the 
\citet{1989ApJ...344..715C,1993ApJ...414..454C} 
and  \PNS\ catalogs, as a function of apparent magnitude.
Filled boxes show PNe identified in the \PNS\ catalog
as members of
NGC~3379, while open stars indicate members of NGC~3384.
}
\label{fig:Ccomp} 
\end{figure}
                  
We have also investigated differences in velocities and magnitudes.
Figure~\ref{fig:Ccomp} shows the correlation between  magnitudes,
and velocity differences, for common objects in the two catalogs. 
The velocities show a median
offset of 28\kms\ and a scatter of 16\kms, with no indication that the
errors increase for fainter PNe. 
  We also looked for systematic
variations in the differences between the results with position in the
field or with velocity itself, which might be expected if the
wavelength calibration of the \PNS\ data contained undiagnosed
systematic problems, but no such effects were apparent.

\subsection{Comparison with Sluis \& Williams (2006)}
\citet{2006AJ....131.2089S} measured the positions, velocities and
magnitudes for a sample of 54 PNe.  Since these results were obtained
using a Fabry--Perot interferometer, they are potentially subject to a
different set of systematic effects, so provide yet another useful
independent check on the reliability of our catalog.  Of their
objects, we recovered 39, with the remainder either masked in our data by star
trails or at small radii where the \PNS\ has trouble recovering objects
against the bright continuum of the galaxy (see \S\ref{sec:comp}).
The  difference in position for the 
matched pairs  is even  smaller than those for the \citet{1989ApJ...344..715C}
set, with the \rms\ combined difference being just  0.9\arcsec.

\begin{figure} 
\includegraphics[width=8.6cm]{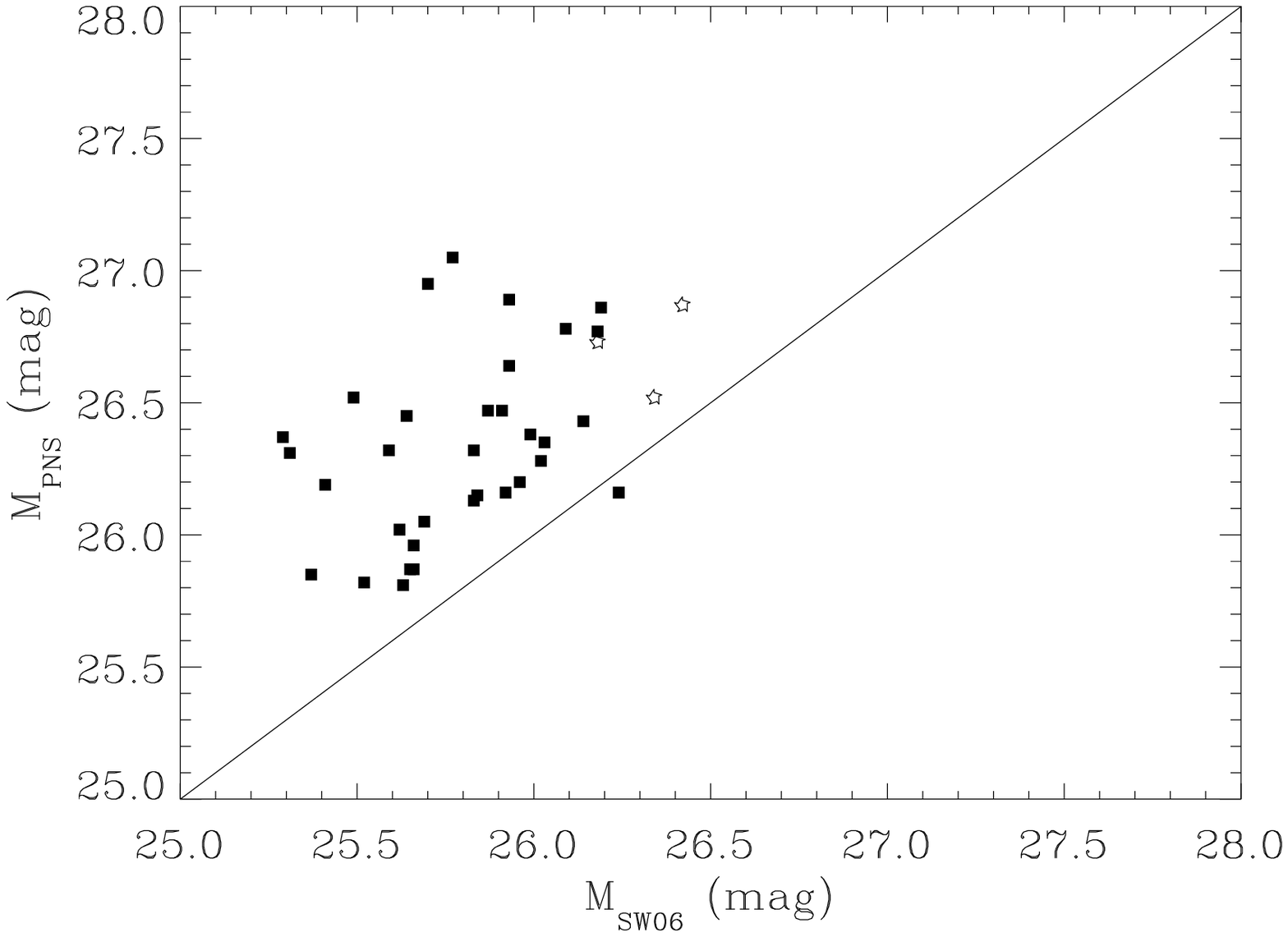}
\includegraphics[width=8.6cm]{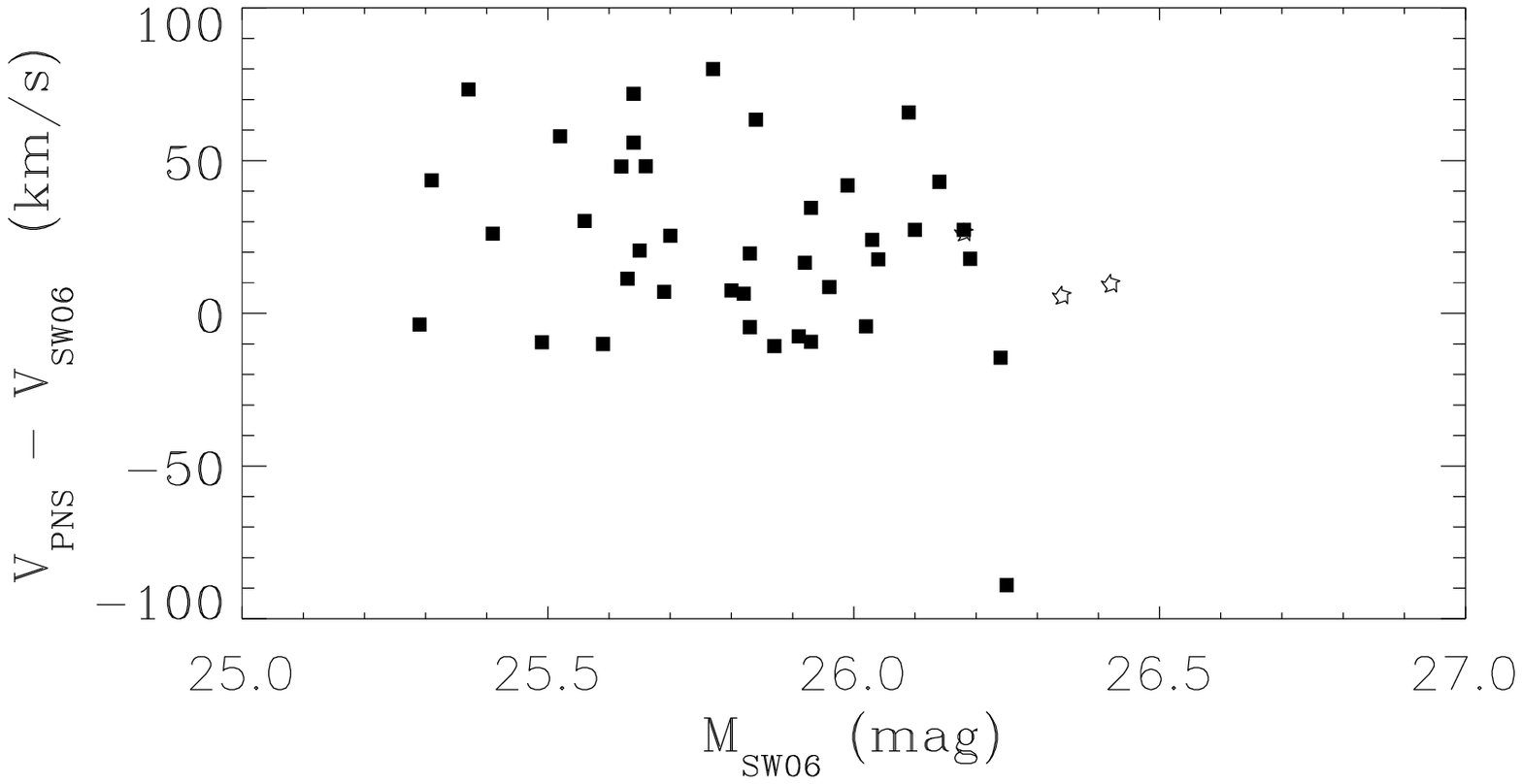} \caption{Comparison 
between the measured magnitudes (upper panel)
and velocities (lower panel) for the PNe common to the 
 \citet{2006AJ....131.2089S} 
and  \PNS\ catalogs, as a function of apparent magnitude.
 Symbols are as in
Figure~\ref{fig:Ccomp}. } \label{fig:Scomp} \end{figure}

\begin{figure}[h] 
\includegraphics[width=8.6cm]{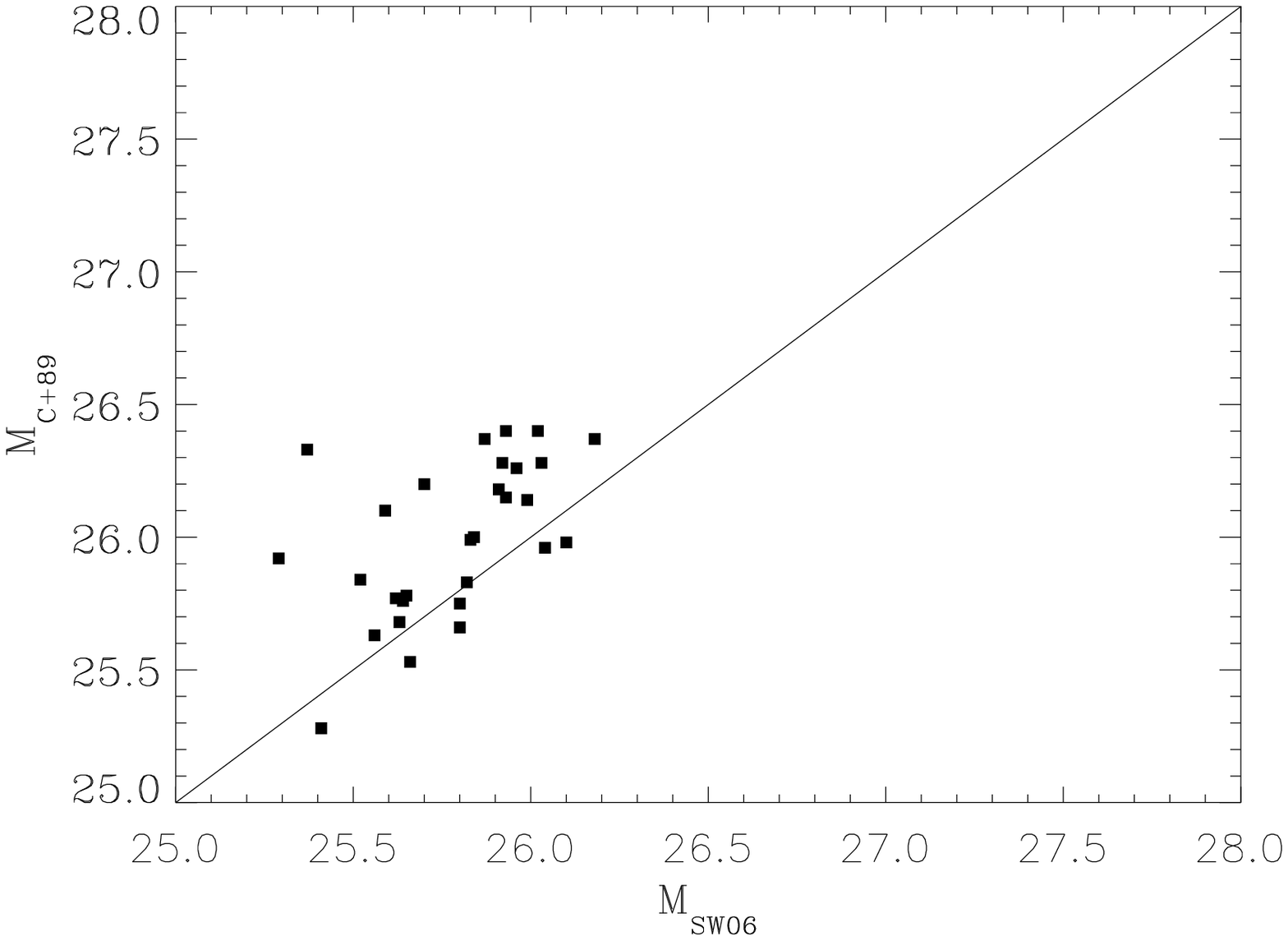}
 \caption{Magnitude comparison between the catalogues of 
\citet{2006AJ....131.2089S} and \citet{1989ApJ...344..715C}
\label{fig:SvCmag}}\end{figure}

Figure~\ref{fig:Scomp} shows the differences between the magnitude
correlation and the velocity differences, between common objects in the two catalogs.  
Once again, the velocities match up
well, with a median offset between the two datasets of only 20\kms, a
combined scatter of 32\kms, and no sign of any
increase in error at faint magnitudes.  
There is again considerable scatter between   magnitudes,
though not much more than is evident between the external catalogues 
themselves (see Figure~\ref{fig:SvCmag}).
Importantly, there are no trends which suggest that the measured velocities are
correlated with magnitude in any way. Indeed  a much more extensive analysis by
\citet{2006MNRAS.369..120M} of the magnitudes of PNe in M31 returned
by the \PNS\ compared to those from other surveys found no such trend,
and we have carried out analogous tests which show that
there is no indication for any systematic errors in velocity or
 magnitude with position in the \PNS\ field.

In summary, internal and external cross-checks have
confirmed that there are no systematic effects that would compromise
our measured velocity dispersions, and that the errors on individual
PN velocities are at an entirely satisfactory level of $\sim 20\,{\rm
km}\,{\rm s}^{-1}$.

\section{Completeness and Contamination of the NGC~3379 PN Catalog}
\label{sec:compcont}
Before this catalog can be used for studying the dynamics of NGC~3379,
it is important that we quantify what fraction of sources we may have
missed, and that we remove any spurious sources that are unrelated to
the target.

\subsection{Completeness}\label{sec:comp}
We already have some sense of the likely completeness of the catalog
by the fraction of previous identifications that we were able to
recover in \S\ref{sec:compare}, which suggested a figure of
about 80\%, including a fairly constant fraction lost in star trails. 
However those comparisons are not adequate to 
determine the losses due to the  diffuse brightness of the galaxy,
which increases to smaller radius,  making the fainter PNe
progressively harder to identify.
To quantify this effect, we have
randomly placed simulated PNe in the data frames, and used the source
searching algorithm of \S\ref{sec:find} to see what fraction are
recovered.  Fainter objects will obviously be harder to find, so we
have populated the planetary nebula luminosity function to different
faintness limits below  $m^*$ (the magnitude of the brightest
PNe given the known PNLF and distance, 
\citep{1989ApJ...339...39J}, here 25.48) to see 
how faint we can go.  
The simulations show that
the detectability of even the brightest ($m^*$) PNe is essentially 
zero within a radius of $20''.$ For PNe with $m^*+1.0$ detectability
rises sharply to a limiting value of 95\% at $R \sim 100''$  radius, while for 
 fainter PNe ($m^*+1.5$)detectability
reaches a limit of 85\% at the same radius. 
 This  nicely illustrates the
complementarity between the study of stellar kinematics using PNe and
those derived from conventional absorption-line spectroscopy: within
one effective radius  of  \Reff\ = 47\arcsec\ 
the diffuse background of the
galaxy makes PNe very difficult to detect, but this same higher
surface brightness makes absorption-line spectroscopy fairly
straightforward.

We can, of course, correct for incompleteness when studying the PNe as
a tracer population, but must be a little careful when doing so.  If,
for example, the PN kinematics varied significantly with their
luminosity, then the observations at small radii, which preferentially
detect the brightest PNe, would not be sampling the same dynamical
population as those at large radii.  Such a phenomenon has apparently
been detected
in NGC~4697 by \citet{2006AJ....131..837S}, but 
was not seen at all in M31 by \citet{2006MNRAS.369..120M}.
Nonetheless, care must be taken to test for such biases in any
detailed dynamical study.

\begin{figure}
\includegraphics[width=8.6cm]{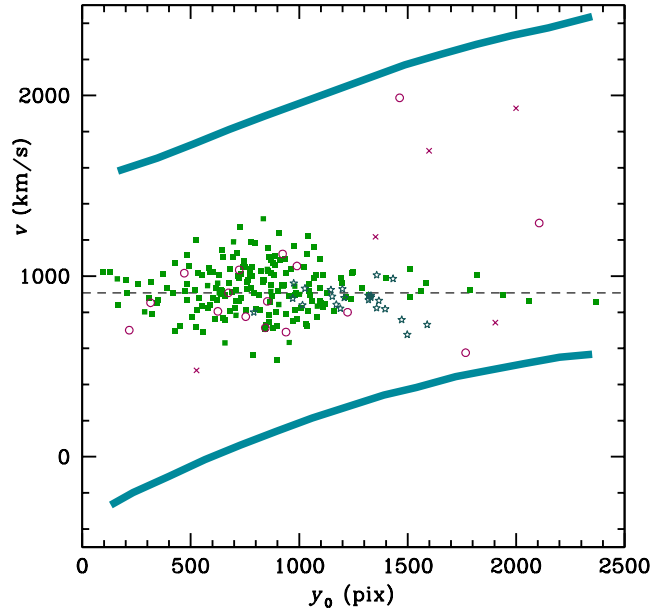}
\caption{Plot of velocity versus spatial coordinate in the 
direction
in which the filter bandpass varies with position due to 
its tilt (here 6\degree).
The points show the detected emission line objects in this
observation: squares are PNe from NGC~3379; stars are PNe from
NGC~3384; circles are extended sources; crosses are unresolved
background sources.  The thick lines mark the detection limits imposed
by the filter FWHM bandpass.} \label{fig:bandpass}
\end{figure}

Perhaps of greater concern is the question of whether the probability
of detection might depend on a PN's velocity as well as its location,
as such a bias would clearly compromise a dynamical study. The 
central wavelength of the narrow-band filter, as used in \PNS, varies 
only slightly over the field, the variation increasing with tilt.
The bandwidth does not vary, to the limit of our measurements.
When the filter has to be tilted the
effects can be mitigated by appropriately
changing the position of the target
in the field of view. A typical situation is shown in
 Figure~\ref{fig:bandpass}, demonstrating how
kinematic bias can be avoided. 

One could imagine other kinematic biases arising from
the few PN pairs which are discarded due to confusion,
or from the impact of the central galaxy light on PN detections.
However, we have performed simulations of these PN ``losses''
and verified that there is no significant velocity selection effect.

\subsection{Contamination by NGC~3384 PNe}\label{sec:n3384}  
Having considered the sources that we may have missed, we now turn to
excess sources that are not  in NGC~3379. 
The SB0 galaxy NGC~3384 lies 
only 7\arcmin{} in projection away from NGC~3379, so their populations 
will appear to some extent superimposed.  NGC~3384 has a 
systemic velocity of 704\kms, some 200\kms\ lower than 
NGC~3379, but inconveniently NGC~3384 is rotating quite rapidly, 
and the side of NGC~3384 closest to NGC~3379 comprises the 
receding part of its disk, which shifts the velocity of any PNe 
to very close to NGC~3379's velocity, so we cannot separate the 
two populations on the basis of their kinematics, as indeed another
glance at  Figure~\ref{fig:bandpass} immediately shows. 

\begin{figure}
\includegraphics[width=8.6cm]{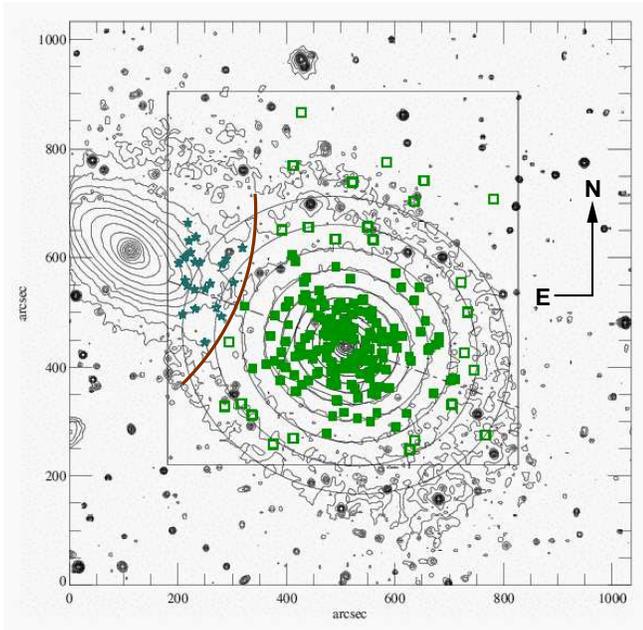}
\caption{$I$-band photometry from \citet{2006MNRAS.366.1126C}, with
the locations of detected PNe and the field of the \PNS\ observation
marked.  
The heavy arc marks the division between the two galaxies
for assigning membership to PNe, so squares are members of NGC~3379
while stars are members of NGC~3384.  The filled symbols for NGC~3379
are used for objects that lie within the isophotal radius out to which
the entire galaxy has been mapped.}
\label{fig:galsep}
\end{figure}

Instead, we adopt the simple approach of dividing the two on a
probabilistic basis: we model the distributions of PNe in the two
galaxies on the assumption that they follow the diffuse starlight, and
assign each PN to whichever distribution makes the dominant
contribution at its location.  One slight subtlety is that because
these galaxies lie at different distances and because their PN
luminosity functions have somewhat different intrinsic specific
frequencies \citep{1989ApJ...344..715C}, the cross-over between the
two does not exactly coincide with the similar point in the raw
photometry.  However, these effects can be accounted for, and the
resulting division between the two, which is shown in  
Figure~\ref{fig:galsep}, allocates 23 of the 214 PNe to NGC~3384.
Although there is some residual
uncertainty in this allocation, its impact is likely to be slight: we
have found that the east-side and west-side kinematics of NGC~3379
remain symmetric even 
if we change the measurement radius  from $200''$
to $240''$.

\subsection{Contamination by non-PNe}\label{sec:cont}
The initial source
identification of \S\ref{sec:find} has already flagged a
number of sources as not being PNe because they are extended either
spatially or spectrally.  Visual inspection of an R-band image
of the field identified many of these as star-forming background galaxies.
At the distance of NGC~3379, even the
largest PNe will be point sources, and the low expansion velocities of
PNe mean that they should also appear unresolved spectrally, so we can
immediately reject such objects.  However, it is interesting to note
that these sources are not uniformly spread spatially and in velocity,
within the limits of the filter bandpass, as one would expect if they
were simply background galaxies with emission lines that happen to
fall within the filter.  As Figure~\ref{fig:bandpass} shows, there is
a definite concentration associated with the position and velocity of
NGC~3379.  Thus there may be a small population of
extended emission-line nebulae associated with this galaxy,
 perhaps similar to the isolated, compact
HII regions discovered in the low-gas density outskirts of other
galaxies \citep{1998ApJ...506L..19F, 2002ApJ...580L.121G, 2004AJ....127.1431R}.
It would
certainly be interesting to study them more closely both
morphologically and spectroscopically with a more conventional
instrument, but for the moment we simply exclude them because they are
not part of the PN population that we are seeking to study
dynamically.

In addition to these resolved objects, there are presumably some 
contaminants that are small and that only contain narrow 
emission lines, and so will not be flagged as extended.  To try 
to identify such sources, we have used the ``friendless'' 
algorithm applied by \citet{2003MNRAS.346L..62M}, in which any 
object that lies more than $n$ standard deviations from the 
velocity distribution of its $N$ nearest neighbors is deemed not 
to be part of the general population due to its kinematic 
peculiarity.  In this case, setting $n=3$ and $N=15$ identifies 
the 5 sources marked as crosses in Figure~\ref{fig:bandpass} as 
likely interlopers; as might be expected, they are spread fairly 
uniformly in position and velocity.  These objects are therefore 
rejected from the default catalog, but we also test our results 
for robustness by seeing how much they change if we only reject 
the two friendless sources that we obtain by setting a more 
conservative $n=5$. This is dealt with in \S~\ref{sec:kins}.

\section{Comparing Stellar and PN Properties}\label{sec:starvPN}
Having produced a clean sample of PNe from NGC~3379, we are now in a
position to compare it to the properties of the unresolved stellar
population.  It is important to ascertain whether they are sampling
the same underlying sources, as we can only exploit the potential
synergy between absorption-line kinematics at small radii and \PNS\
results at large radii if they really form a single tracer
population.  We therefore now compare both
photometric and kinematic properties of PNe to the results available
for the diffuse stellar light to see how well they match up.

\subsection{Spatial Distribution}\label{sec:spatdist}
The simplest comparison that we can make is between the number counts
of PNe and the surface photometry of NGC~3379.  To make a direct
comparison, we must correct the number counts for incompleteness.  As
we have seen in \S\ref{sec:comp}, at small radii PNe are lost
against the bright background of the galaxy, so we use the results
of those simulations
to determine the fraction that
are missing and correct the PN counts accordingly.  At large radii,
the limited field of view of \PNS\ and the presence of NGC~3384
restrict the regions in which PNe can be found, so we must correct for
this geometric factor by using the areas outlined in
Figure~\ref{fig:galsep} to determine the completeness of sampling.

 \begin{figure}
 \includegraphics[width=8.6cm]{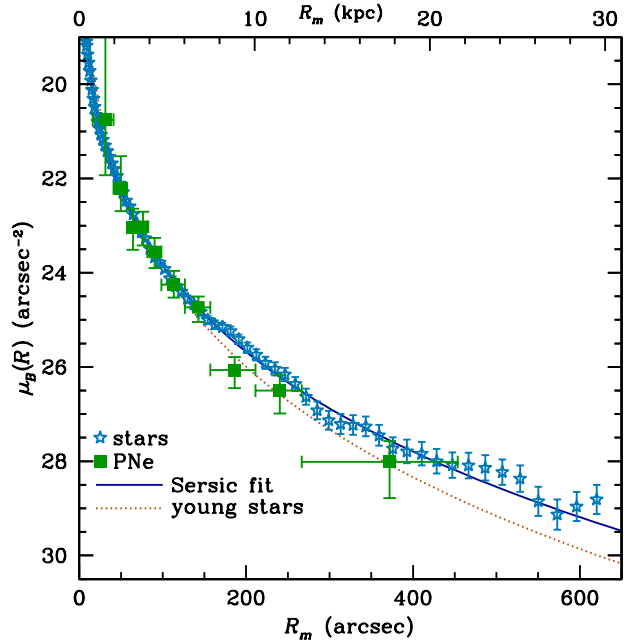} \caption{Radial profiles of
stars and PNe in NGC~3379.  
 Stars show the stellar photometry, and squares  the PN number
counts, with an arbitrary vertical normalization.   The PNe counts have been
corrected for completeness (see \S~\ref{sec:comp}.
Vertical error bars in this and subsequent figures show the 
 1-$\sigma$ uncertainties on the data (based on counting statistics and
 completeness correction uncertainties), while the horizontal bars show
 the radial range over which the PNe have been binned.  The solid line
 is a \citet{1968adga.book.....S} profile fit to the stellar
 photometry, while the dotted line shows the profile for a hypothetical
 younger PN population (see text).}
 \label{fig:spatcomp}
 \end{figure}

For the stellar photometry, we have used the wide-field $B$-band
observations of \citet{1990AJ.....99.1813C}, supplemented by the HST
$V$-band photometry of \citet{2000AJ....119.1157G} to fill in the
photometry within the central 10\arcsec{}.  The photometry has been matched
up by assuming a constant color offset of $B-V=1.03$.  Major axis
profiles have been transformed into intermediate axis values,
$R_m\equiv a \sqrt{1-\epsilon}$, using the ellipticities, $\epsilon$,
from the same literature.  At large and small radii the observational
uncertainties in ellipticity become rather large, so we fix
$\epsilon=0$ for $R_m < 0.8''$, and $\epsilon=0.16$ for $R_m > 40''$.

Figure~\ref{fig:spatcomp} compares the PN number counts (to a limit of
$m^*+1.5$) to the stellar photometry.  Clearly, the two match up
reasonably well, although there might be some indication that the PNe
are more centrally concentrated.  This possibility should be investigated, 
given the suggestion by D+05 that the PN kinematics of
NGC~3379 could be driven by a younger population, which should also be
more centrally concentrated. The figure therefore also
shows a model for the spatial distribution that might be expected for
such a young population, constructed  by fitting a
\citet{1968adga.book.....S} profile to the photometry (also shown in
Figure~\ref{fig:spatcomp}), and multiplying it by an extra $R^{-0.3}$
power law, which approximately matches the difference between the
young and total stellar profiles in the models of D+05.
This model fits the PN number counts  better than the raw stellar
photometry, but no better than a third model (not shown) incorporating
a mild metallicity gradient such as has been seen in other
elliptical galaxies \citep{2005ApJ...627..767M}.

\begin{figure}
\includegraphics[width=8.6cm]{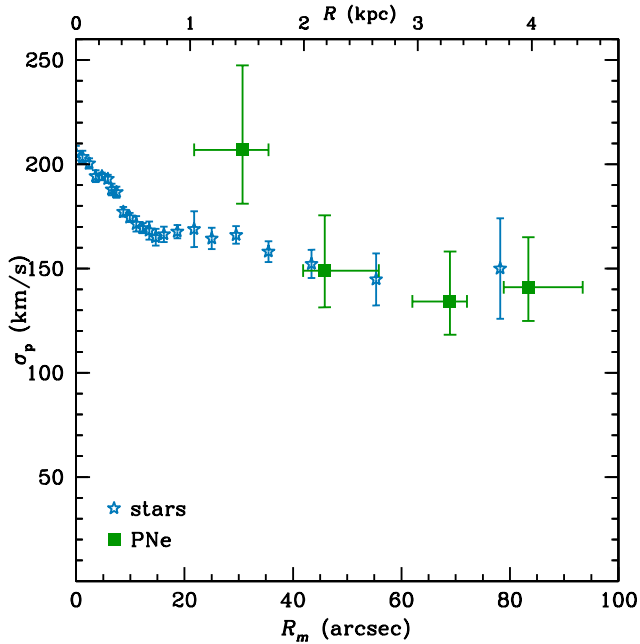}
\caption{Projected velocity dispersion profiles of stars (blue stars) and PNe
(green squares) in the inner parts of NGC~3379.
The stellar data are from \citet{1999AJ....117..839S}.
Horizontal error bars show 68\% of the radial range of the PNe.}
\label{fig:dispcomp}
\end{figure}

\subsection{Kinematics}\label{sec:kins}
Since the primary goal of this project is to map out the stellar
kinematics of elliptical galaxies out to large radii, perhaps the most
important comparison that we can make is between the kinematics
derived from PNe and those from stellar absorption-line data.
Fortunately, \citet{1999AJ....117..839S} obtained deep long-slit
absorption-line spectra of NGC~3379 at 8 different position angles.
It is an indication of the difficulty in obtaining stellar kinematics
out to large radii using conventional methods that even these data
struggle to reach beyond 2~\Reff.  However, as we will see in
\S\ref{sec:vfield}, the kinematics of NGC~3379 are dominated by
random motions, with little evidence for any azimuthal structure.  We
can therefore improve the signal-to-noise ratio without losing much
dynamical information by averaging the \rms\ velocity, $v_{\rm RMS}$,
from the different position angles to form a single stellar velocity
dispersion profile.  Although the mean streaming velocity, $v$, is
much smaller than the velocity dispersion, $\sigma$, at all radii
along all position angles, we retain it in the \rms\ velocity by
defining $v_{\rm RMS} = \sqrt{v^2 + \sigma^2}$.
Figure~\ref{fig:dispcomp} shows the resulting average dispersion
profile, and compares it to the profile derived by binning the PNe in
this range of radii in groups of $\sim 25$ and calculating the \rms\
velocities in the resulting bins.  To within the sizeable
error bars, the kinematics of the PNe are indistinguishable from the
stellar component.

\begin{figure}
\includegraphics[width=8.6cm]{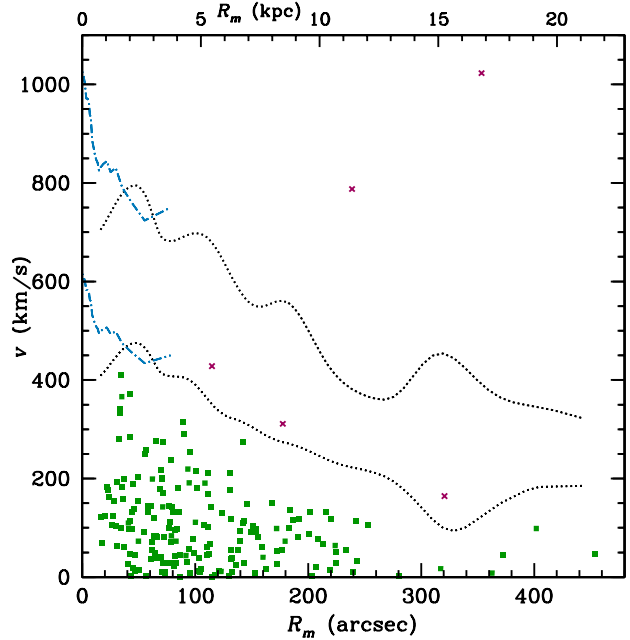}
\caption{Azimuthally-averaged velocity distribution of PNe in
NGC~3379, plotted as a function of their projected radius.
The dotted
lines show the 3-$\sigma$ and 5-$\sigma$ envelopes of their velocity
distribution (and the excluded unresolved point sources from
\S\ref{sec:cont} shown as crosses).  The dash-dotted lines show the
corresponding envelopes derived from the absorption-line stellar kinematics.
A systemic velocity of 906\kms\ has been subtracted.}
\label{fig:pnvdisp}
\end{figure}

The marginally high first point for the PNe in
Figure~\ref{fig:dispcomp} is only $\sim 1.2\sigma$ away from the
stellar data, so is not statistically significant.  Further
reassurance on this point can be derived from
Figure~\ref{fig:pnvdisp}, which shows the full azimuthally-averaged
PNe data set, with lines showing the $3\sigma$ and $5\sigma$ bounds on
the velocity distribution, derived using a moving window to calculate
dispersions over a range of radii, thus creating a smooth profile.
The figure also shows the 3-$\sigma$ and 5-$\sigma$ bounds over the
limited range of radii that are provided by the absorption-line
data.  It is clear that there is no discontinuity between these data
sets, and that the velocity dispersion declines smoothly and
continuously from the absorption-line data at small radii out to the
PNe at large radii.  Since the errors on the
absorption-line kinematics are much smaller than those on the PN data
at small radii, we will not use the PN data from inside $40''$ for the
subsequent analyses.

\section{Kinematic Characterization}\label{sec:vfield}
Having established the reliability of the PNe in NGC~3379 as tracers
of the stars, we now turn to characterizing the global kinematic
properties of the galaxy.  First, we must determine the net velocity
of the galaxy, which must be subtracted for all subsequent analysis of
the system's internal dynamics.  To do so, we simply take the median
velocity of all PNe 
that lie in the region in which the data are
fairly uniformly complete, $60'' < R_m < 200''$ (see
\S\ref{sec:compcont}).  The resulting zero-point velocity is found to
be $906\,{\rm km}\,{\rm s}^{-1}$, in surprisingly good agreement with
the established value of $911\pm2\,{\rm km}\,{\rm s}^{-1}$
\citep{2000MNRAS.313..469S} given the potential for residual
systematic velocity errors in the \PNS\ data (see \S\ref{sec:compare}).
In order to minimize the impact of any systematic offset in the \PNS\
results, we adopt the $906\,{\rm km}\,{\rm s}^{-1}$ as the zero point
for these data, although it really is not  critical, since
changes in this value of up to $20\,{\rm km}\,{\rm s}^{-1}$ have no
significant impact on the derived dynamics.

\subsection{Two-Dimensional Velocity Field}
Before we can reduce the kinematics of NGC~3379 to more tractable
one-dimensional profiles, it is important to ascertain that the
dynamics of the system are sufficiently simple that these
one-dimensional functions capture the essence of the object's
kinematics.  We therefore begin by taking a more general look at the
full velocity field.

\begin{figure}
\includegraphics[width=12cm,viewport=90 0 540 305]{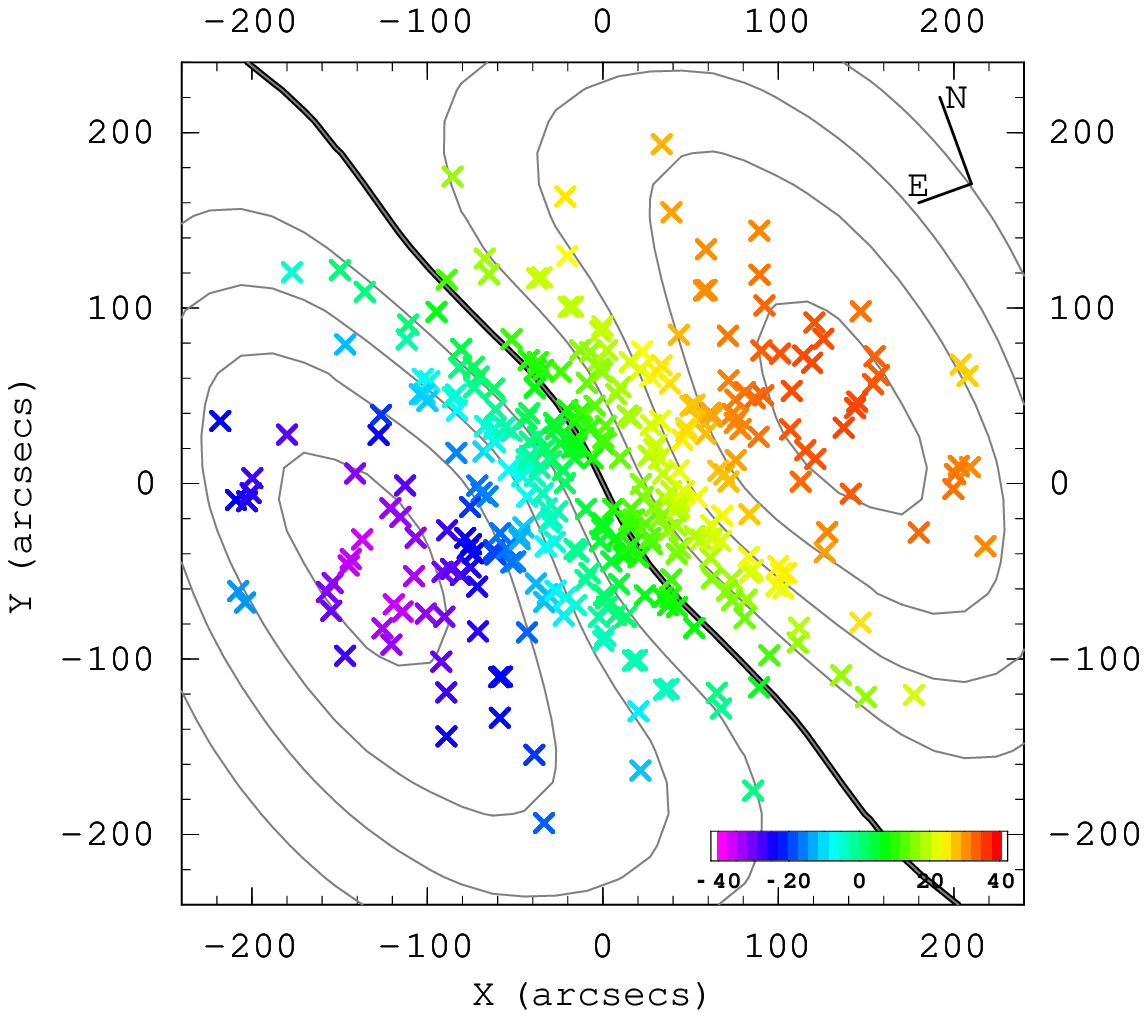}
\includegraphics[width=12cm,viewport=90 0 540 320]{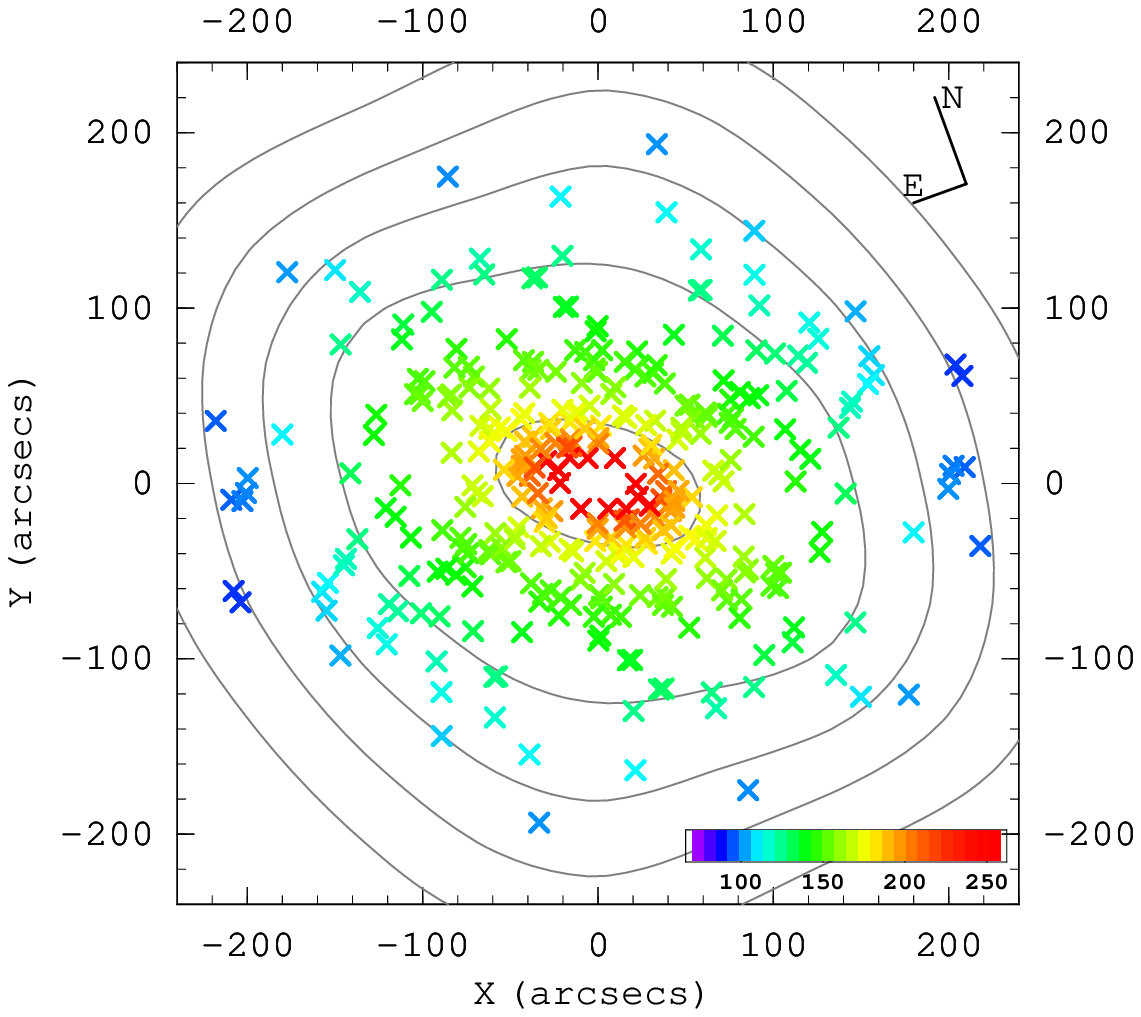}
\caption{Smoothed velocity field of PNe in NGC~3379,
  with the photometric major axis (PA=70$^\circ$) oriented in the
  $x$-direction.  Upper: Mean velocity field, with individual PN
  velocities plotted as crosses: see shaded bar for velocity scale.
  Iso-velocity
  contours are also plotted, with the heavy contour at zero
  velocity and subsequent contours at 10 \kms\ intervals.  
   Lower: Dispersion (\rms\ velocity) field. Again, the bar shows
  the  scale, and the contours run inwards from 60 \kms\ to 180 \kms  }
\label{fig:pnsmooth}
\end{figure}

Of course the kinematic detail which we can reliably recover is
limited by the finite statistics of the PN data, but this limitation
can be mitigated by some reasonable assumptions, using an approach
patterned after \citet{2004ApJ...602..685P}.  The first assumption
that we adopt is that the system has at least a triaxial symmetry,
which implies that each position in phase space $(x,y,v)$ has a
``mirrored'' counterpart $(-x,-y,-v)$. We thus map all the PN data
through this transformation and thereby double the sample size used to
recover the velocity field.  We also assume that the velocity field
is relatively smooth and that the strongest observed inhomogeneities
are due to small-number statistics. We therefore apply a smoothing
function to the PN data to accentuate the true underlying velocity
field. We first use a median filter in order to improve the
signal-to-noise ratio per spatial bin, then smooth by convolving with
a Gaussian kernel with width $\sigma \sim46''$ (2.2 kpc). 
We replace each PN
velocity by the local smoothed value, and also calculate the mean
velocity and dispersion everywhere in the field of view.  The
resulting smoothed mean velocity and velocity dispersion fields are
shown in Figure~\ref{fig:pnsmooth}.

\subsection{Rotation}
 The most dramatic feature in Figure~\ref{fig:pnsmooth} is the clear
rotation of the system, which is non-zero at a significance level of
3-$\sigma$.  The projected rotation axis lies at a position angle of
$\phi_0 \sim 20^\circ$ east from north.  This does not align with
either the photometric major axis ($\sim 70^\circ$) or the photometric
minor axis ($\sim -20^\circ$), implying
that the galaxy cannot be a simple axisymmetric system.
We note that the precise value of
$\phi_0$ from our data is somewhat uncertain, because systematic
offsets in our PN velocities are of order $\sim 20$ km/s, about half
of the rotation amplitude in Figure~\ref{fig:pnsmooth}. However,
further indication of the galaxy's triaxial nature comes from matching
these kinematics to the absorption-line data at small radii, where the
projected rotation axis is found to be $\sim -15^\circ$
\citep{1999AJ....117..839S}; such a twist in the rotation axis cannot
occur in an axisymmetric system.

To better quantify this rotation, we have fitted the raw PN velocities
with a simple rotation curve model, 
\begin{equation}
v(\phi) =v_0 \times \sin (\phi-\phi_0),
\end{equation}
where $\phi$ is the position angle of the PN on the sky as measured
from the center of NGC~3379, and the constants $v_0$ and $\phi_0$
respectively measure the amplitude of rotation and its projected axis.
This fitting procedure has the advantage of being insensitive to the
azimuthal completeness of the data; its applicability is discussed in
\citet{2001A&A...377..784N}.  The resulting best-fit parameters are
$v_0 = 45\pm 15\,{\rm km}\,{\rm s}^{-1}$ and $\phi_0 = 20^\circ\pm20^\circ$.
We
also looked for any indications of variation in these parameters with
radius by dividing the data into three radial bins, but to within the
errors the data are consistent with a flat rotation profile at a fixed
position angle.  

\begin{figure} 
\includegraphics[width=8.6cm]{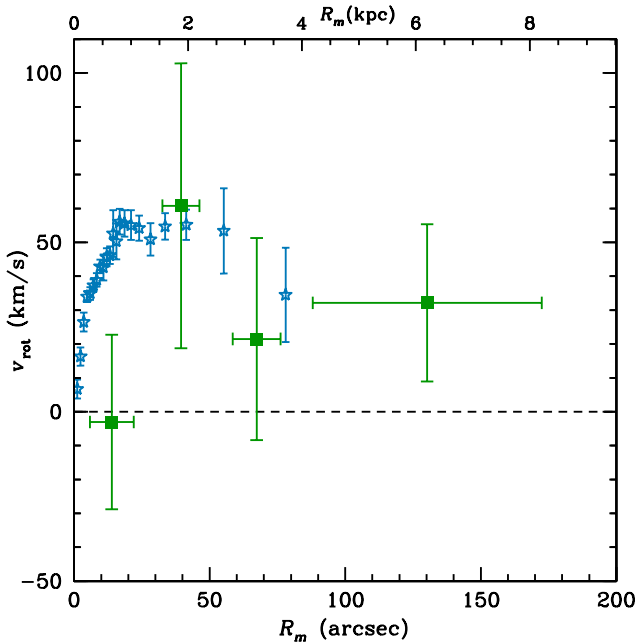}
\caption{ Mean rotational velocity of stars and PNe along the major
axis of NGC~3379.  Boxes show \PNS\ data and stars show
absorption-line kinematics from \citet{1999AJ....117..839S}.  }
\label{fig:pnrot} 
\end{figure}
We can also construct a more conventional major-axis rotation profile,
for direct comparison with the major-axis long-slit absorption-line
data.  Figure~\ref{fig:pnrot} shows the results obtained by combining
the PN data from a strip 50\arcsec{} wide aligned with the photometric
major axis, and mirrored through the center of the galaxy to double
the sample size.  The figure also shows the corresponding
absorption-line data from \citet{1999AJ....117..839S}: in the central
bin, the PN velocities are suppressed because the wide strip samples a
long way up the minor axis of the galaxy, but at larger radii the
amplitude of rotation is entirely consistent between \PNS\ and
absorption-line data.  It is also interesting to note that these data
show no indication of the high rotation velocity predicted in the
outer parts of elliptical galaxies by merger scenarios
\citep{1996ApJ...460..101W}, although in any individual case such as
this it is always possible that most of the rotation lies undetectably
in the plane of the sky.

\subsection{Velocity Dispersion Profile}
\label{sec:dispsec}
It is apparent from inspection of Figure~\ref{fig:pnsmooth} that
random motions dominate over the mean streaming motions throughout
NGC~3379, so by accurately characterizing the random velocities we
will capture most of the dynamical information that these data
contain.  Further, it is clear from Figure~\ref{fig:pnsmooth} that the
velocity dispersion field is regular in appearance, and roughly
aligned with the stellar isophotes.  We can therefore characterize
these random motions quite accurately with a simple
azimuthally-averaged one-dimensional profile as a function of the
intermediate axis radius, $R_m$.

\begin{figure}
\includegraphics[width=8.6cm]{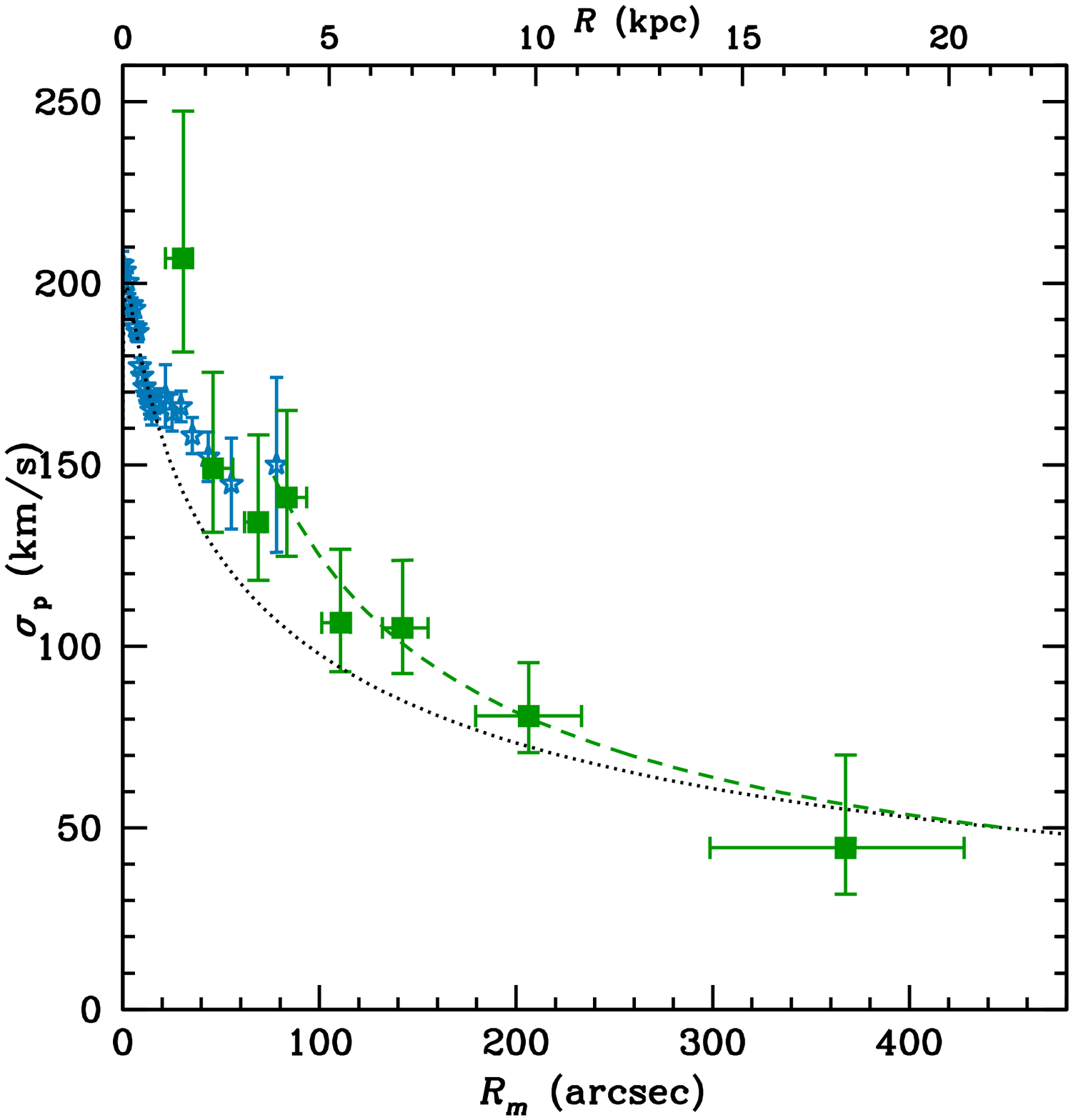}    
\includegraphics[width=8.6cm]{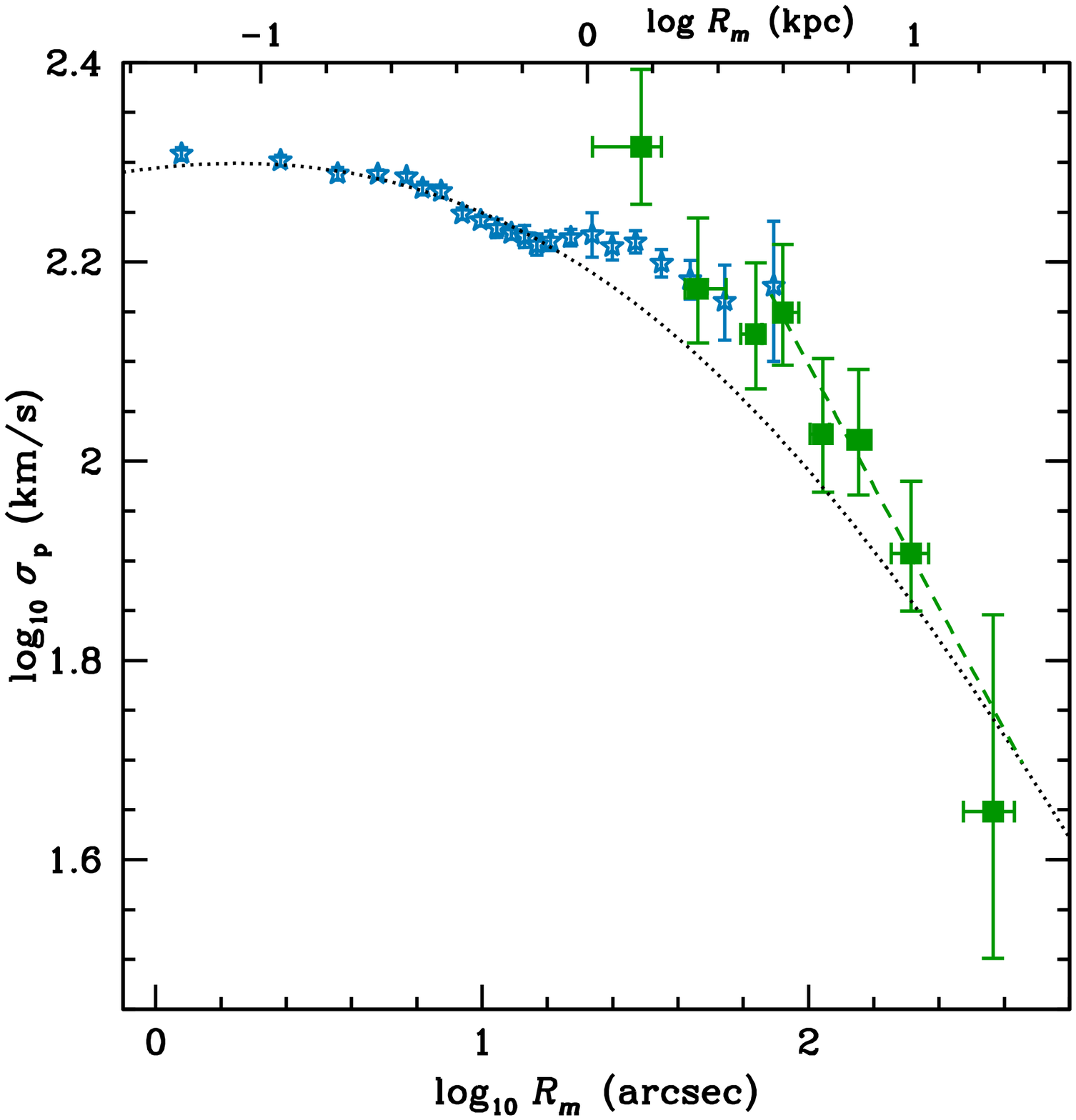}    
\caption{Projected velocity dispersion profile for NGC~3379, showing
data from stars (stars) and PNe (boxes), plotted on both linear (upper figure) and
logarithmic axes (lower figure).  
Dashed lines show a power-law fit  
at large radii, while dotted lines
show an isotropic constant mass-to-light ratio model.  }
\label{fig:dispprof}
\end{figure}

As described in \S\ref{sec:kins}, we have calculated the empirical
binned PN velocity dispersion profile, which we combine with the
stellar absorption line data of \citet{1999AJ....117..839S},
presenting the result in Figure~\ref{fig:dispprof}.  The most striking
feature of this plot is the rapid decline in the velocity dispersion
with radius.  A maximum-likelihood fit to the unbinned data at radii
$ R_m > 80''$ (see below) gives a best-fit power law with an index of $-0.6
\pm 0.2$.  A possible concern is that the profile might be
sensitive to the manner in which PNe have been flagged as
contaminating objects in the tails of the distribution by the
friendless algorithm (see \S\ref{sec:cont}): if the velocity
distribution is significantly non-Gaussian at large radii, this
algorithm could be overzealous in trimming objects from the tails of
the distribution, systematically underestimating the true velocity
dispersion.  We have investigated this effect by constructing the
dispersion profile after applying a
more conservative application of the friendless algorithm that only
truncates the velocity distribution at $5\sigma$, but find the
difference to be insignificant. Thus, the decline in velocity dispersion with radius
does not appear to be an artifact arising from the assignment of
galaxy membership.
In what follows we will apply 
the more usual $3\sigma$ cut, and  PNe lying beyond this cut
will no longer be plotted.
This proximity in the slope of the velocity dispersion profile to the
Keplerian value of $-0.5$ has been seen repeatedly in earlier PN data for
NGC~3379 (\cite{1993ApJ...414..454C}, R+03,
\cite{2006AJ....131.2089S}), but is extended by the new data set
to even larger radii.  Since at these radii we are well outside most
of the luminous component of the galaxy, it acts very much like a
point-mass potential, so the simplest interpretation of the Keplerian
decline is that the mass is dominated by the luminous component and
that we are seeing close-to-Keplerian isotropic orbits in its
potential.  However, as discussed in the Introduction, there is always
the possibility that a more complex orbit distribution could mimic
this behavior in a different gravitational potential.  Indeed, it is
apparent from Figure~\ref{fig:dispprof} that there is a fair amount of
structure in the velocity dispersion profile, including a rapid
transition from a shallow inner slope  
to a steep outer slope at a
radius of $\sim 80''$.  Due to the long-range nature of the
gravitational force, such rapid changes are difficult to induce by
varying the gravitational potential, which at least suggests that a
transition in the orbital structure of the galaxy is occurring.

\subsection{Higher-Order Velocity Moments}\label{sec:higher}
One clue as to the true nature of the orbital structure is provided by 
higher order velocity moments.  In particular, the fourth moment, usually 
presented as the non-dimensional kurtosis, $\kappa \equiv \langle v^4 
\rangle /\langle v^2\rangle ^2 -3$, offers a useful diagnostic.  In 
systems where the orbits are preferentially circular, the observed 
line-of-sight velocities tend to pile up at the circular speed, in extreme 
cases producing a line-of-sight velocity distribution with two peaks at 
the positive and negative circular speed.  Such distributions are 
characterized by negative values of $\kappa$.  Conversely, a system 
containing objects on radial orbits will tend to produce a centrally 
peaked line-of-sight velocity distribution with long tails, resulting in a 
positive value of $\kappa$. The halo of a galaxy with isotropic orbits and 
a constant circular-velocity curve (i.e. with a massive dark-matter halo) 
has a Gaussian line-of-sight velocity distribution, and $\kappa=0$ 
\citep{1993MNRAS.265..213G}.

\begin{figure}
\includegraphics[width=8.6cm]{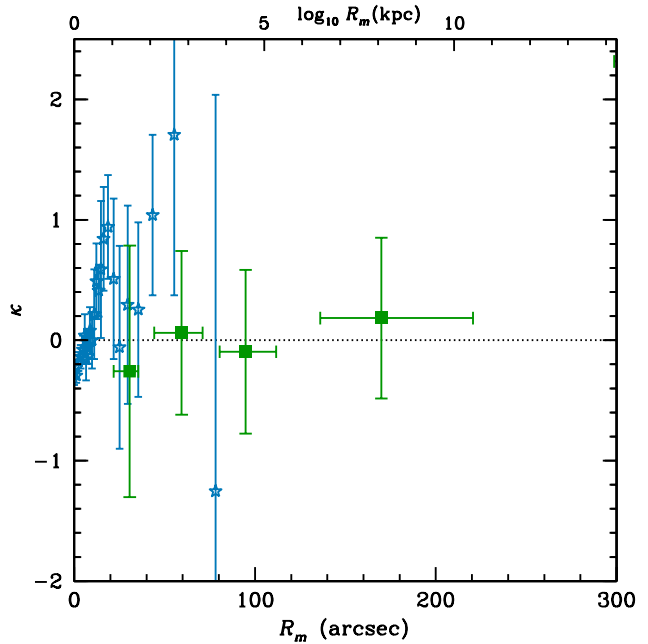}
\caption{
NGC~3379 projected kurtosis profile, showing data from stars
(stars) and PNe (squares).
}
\label{fig:kurt}
\end{figure}
We have therefore calculated the $\kappa$ profile for the PNe in NGC~3379, 
with the results presented in Figure~\ref{fig:kurt}.  The figure also 
shows the higher-order moments obtained from the absorption-line spectra 
of \citet{1999AJ....117..839S}: they obtained the related Gauss-Hermite 
coefficient $h_4$ \citep{1993ApJ...407..525V,1993MNRAS.265..213G}, which 
we have transformed into the kurtosis by the approximate relation $\kappa 
\simeq 8 \sqrt{6} h_4$.

The {\it stellar} kurtosis in Figure~\ref{fig:kurt} shows significant departures 
from zero, which may reflect structural transition regions in the galaxy's 
central parts, or systematic observational uncertainties 
\citep{1994MNRAS.269..785B}. We are most concerned with the kurtosis of 
the PNe in the halo, where we might see a signature of radially-biased 
orbits (see \S\ref{sec:intro}). As seen in Figure~\ref{fig:kurt}, here the 
kurtosis is consistent with zero out to $\sim 200''$ or
$\sim 4$\Reff\ (outside this radius, the small number of
data points make the kurtosis measurement unstable). The PN data therefore 
rule out the more pathological models of extreme radial or tangential 
orbit distributions, but the constraint is still too weak to directly 
preclude a massive dark matter halo with mild radial anisotropy.  
Stronger constraints can be 
made with sophisticated full orbit modeling techniques as undertaken by 
R+03, incorporating the information from the shape of the velocity 
distribution in a more robust way than is possible with binned moments.

\section{Dynamical Models}\label{sec:dynamics}
A full analysis of the dynamics of NGC~3379 is beyond the scope of
this paper, but we can obtain some insights, and hopefully move toward
resolving the issue regarding the presence or otherwise of a dark
halo, by a fairly simple spherically-symmetric Jeans-based analysis.

\subsection{Is A Spherical Model Reasonable?}
The assumption of spherical symmetry does not seem an unreasonable
proposition given NGC~3379's approximately circular appearance, and we
can always modestly 
transform projected coordinates into an exactly
circularly-symmetric approximation by mapping all quantities along
isophote contours to the intermediate axis radius, $R_m \equiv
\sqrt{ab} = a\sqrt{1-\epsilon}$.  However, there is of course no
guarantee that a roughly circular object is intrinsically close to
spherical, since it could be axisymmetric and viewed along the axis of
symmetry, or even have a more complex three-dimensional shape.

Even with this imponderable, we can make a probabilistic estimate
of the galaxy's three-dimensional stellar ellipticity, $\epsilon_\rho$,
based on the known properties of the general population of elliptical
galaxies.  Given an observed projected axis ratio $q_I$, we can find
the probability of the intrinsic short axis ratio $q_\rho$ via:
\begin{equation}
P(q_\rho |q_I) \propto P(q_\rho)P(q_I|q_\rho),
\end{equation}
where the distribution of galaxies' intrinsic axis ratios,
$P(q_\rho)$, has been estimated by \citet{1992MNRAS.258..404L}.
If we assume an oblate galaxy, then we also know that
\begin{equation}
P(q_I|q_\rho) \propto \frac{q_I}{\sqrt{(1-q_\rho^2)(q_I^2-q_\rho^2)}}
\end{equation}
\citep[][equations 4.30 and 4.32]{1998gaas.book.....B}, and so can
solve for the posterior probability distribution.  Given the observed
value $q_I=0.84$ for NGC~3379, the median value of the resulting
axis-ratio distribution is $q_\rho \approx 0.7$.  If we relax the
assumption of oblateness to a more realistic triaxial distribution,
the median $q_\rho$ becomes even larger, so we can say that it is
odds-on that the intrinsic ellipticity of the NGC~3379 stellar density
distribution is less than 0.3.  Assuming any additional mass
contributed by dark matter is at least this close to spherical,
this measurement then places an upper limit on the ellipticity of the
total density.  Further, the gravitational potential is always rounder
than the mass distribution that generates it, so in this case we
obtain an upper limit on the potential axis ratio of $\sim 0.1$ and we
expect the spherical approximation to be good to $\sim$10\%
\citep{2000A&AS..144...53K}.  Of course this
calculation is probabilistic in nature, and we could just be unlucky
to find a highly aspherical galaxy projected as round on the sky, but
we have at least demonstrated that a spherical model is a plausible
place to start.

We therefore proceed to analyze the dynamics of this system assuming
spherical symmetry.  Under this assumption, the Jeans equation
relating velocity dispersion to the mass distribution can be written in the
compact form
\begin{equation} \label{eq:Jeanseqn} 
{G M(r) \over r} 
  = v_{\rm c}^2(r) = \left[\alpha(r)+\gamma(r)-2\beta(r)\right]\sigma_r^2(r), 
\end{equation} 
where $M(r)$ is the mass contained within radius $r$, which can be
transformed into the circular speed $v_{\rm c}$.  On the right-hand
side, $\alpha$ is the logarithmic gradient of the tracer population density $j(r)$,
(here the PNe),
$\alpha\equiv -d\ln j/d\ln r$, $\gamma$ is the logarithmic gradient of
the radial mean-square velocity, $\gamma\equiv -d\ln \sigma_r^2/d\ln
r$, and $\beta$ is the usual anisotropy parameter, 
\begin{equation}\label{eq:beta}
\beta \equiv 1 - \sigma^2_{\theta}/\sigma^2_r,
\end{equation}
defined such that isotropic systems have $\beta=0$, radially-biased
systems have $\beta > 0$, and tangential bias gives $\beta<0$. The
tangential mean-square velocity, $\sigma^2_\theta$, contains both the
mean streaming and random component, as set out in
\S\ref{sec:kins}.  However, since we saw in
\S\ref{sec:vfield} that the random motions dominate everywhere,
it can loosely be thought of as just the random component.

Although equation~\ref{eq:Jeanseqn} looks simple enough, the large
number of unknown functions means that it is far from trivial to
solve.  The only quantity that we can disentangle from the dynamical
terms is $\alpha(r)$, since this is given fairly directly by the
observed spatial distribution of the tracer population.  We therefore
deal with this term first, before returning to the more general
solution of equation~\ref{eq:Jeanseqn}.

\subsection{The Spatial Distribution 
of the Kinematic Tracer}\label{sec:stellarmod} 
           As we saw in Figure~\ref{fig:spatcomp}, the projected light
distribution of both stars and PNe is well represented by a
\citet{1968adga.book.....S} law fit to the surface brightness
(measured in magnitudes per square arcsecond$^{-2}$),

\begin{equation}\label{eq:sersic}
\mu(R) -\mu(0) \propto (R/a_s)^{1/m}.  
\end{equation}
The best-fit parameters for this system are $m=4.74$, $a_S =
0.0013''$, and a central surface brightness of $\mu_0 = 12.2\,{\rm
mag}\,{\rm arcsec}^{-2}$ in the $B$-band.
This fit breaks down in the
innermost arcsecond of NGC~3379, but it is entirely adequate for the
large-scale modeling purposes of this analysis  
(see Figure~\ref{fig:spatcomp}).
After allowing for a 
Galactic extinction of $A_B=0.105$ \citep{1998ApJ...500..525S}, the
model fit integrates to yield a total absolute magnitude of 
$M_B=-19.8$.  The scale parameter can also be transformed into an
estimate of the effective radius
 of $R_{\rm S,eff} = 47''$, a value which is used throughout this paper.
As an aside, we note that this value does not match very well with the more
usually adopted value of \Reff = 35.5\arcsec\ 
\citep{1990AJ....100.1091P} 
but also that none of the analysis here depends on the value
adopted.   
The observed light distribution of
equation~\ref{eq:sersic} deprojects to a luminosity density of
\begin{equation}
j(r) \propto (r/a_{\rm S})^{-p} \exp{[-(r/a_{\rm S})^{1/m}]},
\end{equation}
where $p$ is a known function of $m$ \citep{1997A&A...321..111P}, and
we can then readily calculate the logarithmic radial gradient of this
quantity as a function of radius, $\alpha(r)$, or integrate it out in
radius to obtain the luminosity contained within radius $r$, $L(r)$.
This quantity can then be compared to the enclosed mass at any radius
via the mass-to-light ratio, $\Upsilon(<r) = M(r)/L(r)$.  

We can also calculate the contribution from the stars themselves to
$M(r)$ if we know the mass-to-light ratio of the stellar population,
$\Upsilon_*$, which allows us to convert the luminosity density into
the stars' contribution to the total mass density, $\rho_*(r) =
\Upsilon_* \times j(r)$.  The appropriate value for the stellar
mass-to-light ratio of NGC~3379 has been studied in some detail by
\citet{2006MNRAS.366.1126C}, who used some of the latest stellar
population synthesis models to match the observed spectrum of the
galaxy to a credible stellar population, and hence infer its mass.
From this analysis, the conversion from $B$-band luminosity 
density into stellar mass density is
\begin{equation}\label{eq:stellarML}
\Upsilon_{*,B}=(7.3\pm0.7)\Upsilon_{\odot,B},
\end{equation}
where $\Upsilon_{\odot,B}$ is the mass-to-light ratio for the Sun in
this band.  One uncertainty in any such calculation is the form of the
stellar mass function, since it is always possible to hide mass in the
form of copious low-mass stars that contribute little to the total
luminosity.  The calculation by \citet{2006MNRAS.366.1126C} was based
on the current best estimate for this mass function from
\citet{2001MNRAS.322..231K}, but it is worth noting that the adoption
of the traditional \citet{1955ApJ...121..161S} mass function raises
the mass-to-light ratio by $\sim 30\%$, so the value in
equation~\ref{eq:stellarML} is if anything a lower bound on the
stellar mass-to-light ratio.

\subsection{Solving the Jeans Equation}
\label{sec:Jeans}
At this point, there are several ways of confronting
equation~\ref{eq:Jeanseqn} with the kinematic data to try to infer
something about the dynamics of the tracer population and the
potential within which it orbits.  The most straightforward approach
is to assume a model for both the mass distribution $M(r)$ and the
anisotropy of the orbits, $\beta(r)$, and solve the equation for
$\sigma(r)$.  Both components of the velocity dispersion are then
specified through equation~\ref{eq:beta}, and we can project these
dynamics back onto the sky for comparison with the observed velocity
dispersion profile.

Given the near-Keplerian decline discussed in \S\ref{sec:dispsec}, and
the extensive argument over whether or not NGC~3379 has a massive
halo, the obvious place to start is with the simplest plausible model
in which the mass follows the tracer population and the orbits are
isotropic ($\beta \equiv 0$).  There is only one free parameter in
this model, the constant mass-to-light ratio $\Upsilon$, which we can
determine by normalizing the predicted velocity dispersion profile to
the observed one.  Since this model is most credible as an
approximation at small radii, where everyone agrees that the luminous
component is more likely to dominate the mass, we have carried out
this normalization over of the range $2'' < R < 15''$.  The resulting model
line-of-sight velocity dispersion profile (making use of 
formulae in M{\L}05) is reproduced in
Figure~\ref{fig:dispprof}.  In the central region, the fit to the
observed velocity dispersion is very good, suggesting that this model
is a satisfactory description of the inner parts of NGC~3379; 
indeed,
previous analyses of the kinematics of the central part of NGC~3379
have consistently found that it can be modeled in this simple way
\citep{2000A&AS..144...53K, 2000AJ....119.1157G, 2006MNRAS.366.1126C}.
The normalization implies a central mass-to-light ratio relative to Solar in
the $B$-band of $\Upsilon_{B}=6.3\Upsilon_{\odot,B}$.  Comparing this
value to the contribution from the stellar component,
equation~\ref{eq:stellarML}, it is clear that most, if not all, of the
mass at these radii can be attributed to the stars; if we want clear
evidence of dark matter, we will have to look further out in the
galaxy.

Indeed, at larger radii in Figure~\ref{fig:dispprof} it is apparent
that this simple model and the velocity dispersion data differ,
due to the changes in slope in the velocity dispersion profile
commented upon in \S\ref{sec:dispsec}.  To try to understand the
significance of this structure in the profile, we next consider a
simple phenomenological model for the intrinsic velocity dispersion
profile,
\begin{equation} 
\sigma_r = \sigma_0 \left(\frac{r}{r_0}\right)^{-\delta}
                   \left[1+\left(\frac{r}{r_0}\right)^\eta \right]^{-1}. 
\label{eq:v0eqn} 
\end{equation} 
This model is not motivated by any prejudice as to the mass
distribution, but simply as an attempt to reproduce the observed
change in logarithmic slope of the velocity dispersion at large radii.
Once we have adopted a functional form for $\beta(r)$, the velocity
dispersion is fully specified, and we can project the model for
comparison with the line-of-sight velocity dispersion data.  By
iteratively adjusting the free parameters in equation~\ref{eq:v0eqn},
$\{\sigma_0,r_0,\eta,\delta\}$, the model can then be fitted to the
observed profile.  Since we are primarily interested in explaining
the large-scale change in the slope of this profile, and not the
detailed structure at intermediate radii, we only fit to the data at
$R < 8''$ and at $R > 40''$.  Once we have such a model that fits the
kinematic data, we can solve equation~\ref{eq:Jeanseqn} for $M(r)$,
and hence calculate the enclosed mass-to-light ratio as a function of
radius.  

\begin{figure}
\includegraphics[width=8.6cm]{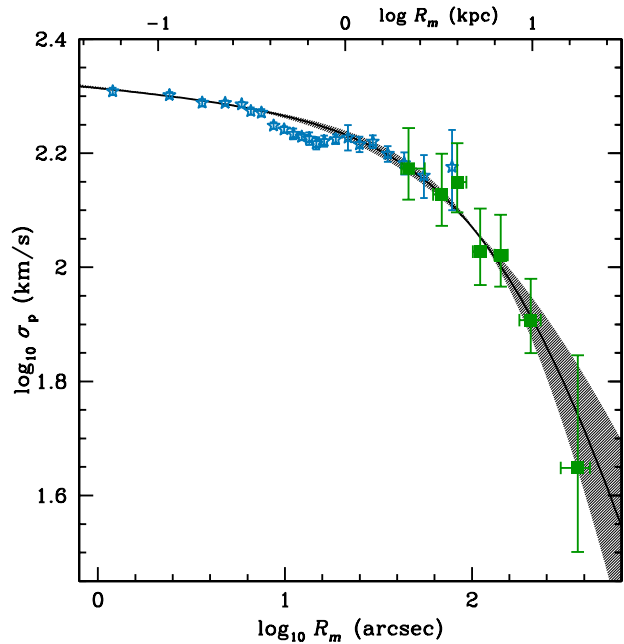}
\caption{Phenomenological fit to the NGC~3379 velocity dispersion data using
the model of equation~\ref{eq:v0eqn} assuming $\beta \equiv 0$.  The
line shows the best fit and the shaded region shows the range covered
by acceptable fits.}
\label{fig:modelfitsone}
\end{figure}
Again, we start with the simplest possibility of an isotropic galaxy
with $\beta \equiv 0$.  As Figure~\ref{fig:modelfitsone} shows, the
model can reproduce the observed velocity dispersion data very well over
the fitted radial range. 
Obviously the data do not uniquely specify the 
model parameters, but, as the shaded region shows, the acceptable fits
capture the range of possible smoothly-varying dispersion profiles
fairly comprehensively.  

\begin{figure}
\includegraphics[width=8.6cm]{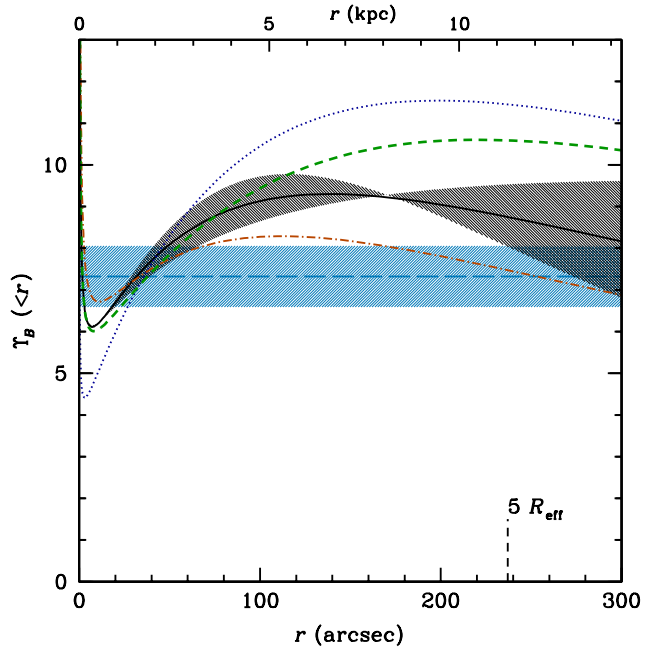}
\caption{Cumulative mass-to-light ratio profiles obtained by solving
the Jeans equation using the phenomenological velocity dispersion
model of equation~\ref{eq:v0eqn}.  The solid line and shaded region
show the best fit and the acceptable range of values derived from the
fits in Figure~\ref{fig:modelfitsone} for an isotropic model; the
dotted line shows the best-fit radially-biased model; the dot-dash
line shows the best-fit tangentially-biased model; and the short-dashed line
shows the best-fit variable-anisotropy model. The long-dashed
line and surrounding shaded error region show the stellar mass-to-light ratio. }
\label{fig:modelfitstwo}
\end{figure}
The result of inserting this model in equation~\ref{eq:Jeanseqn} and
solving for $\Upsilon_B(<r)$ is shown in
Figure~\ref{fig:modelfitstwo}.  The figure also shows the
mass-to-light ratio of the stellar population from
equation~\ref{eq:stellarML}.  As qualitatively predicted, the enhanced
velocity dispersion at large radii translates into an increasing
mass-to-light ratio, but the effect is far from dramatic.  The PN data
become very sparse outside $240''$ (5\Reff), so if we adopt this
as a benchmark value, we find that the mass-to-light ratio within this
radius is $\Upsilon_{{\rm 5}, B} =
8.7^{+0.8}_{-0.7}$~$\Upsilon_{\odot,B}$.  Given the value for the
stellar component from equation~\ref{eq:stellarML}, we thus infer a
dark matter fraction within this radius of between 5\% and 30\%.

To this point, we have imposed the constraint of isotropy on the
orbits, and clearly if we relax this imposition then the limits on the
dark matter fraction will become even weaker.  Simulations and the
weak limits imposed by the higher order moments discussed in
\S\ref{sec:higher} suggest that plausible limits on the anisotropy
parameter are $-0.5 < \beta < 0.5$, so as a simple first test of
relaxing the orbital constraint we have repeated the above analysis
using the extrema of this distribution.  As the famous degeneracy
between $\beta(r)$ and $M(r)$ would imply, models with $\beta=\pm0.5$
can reproduce the observed velocity dispersion profile with the same
accuracy that we saw for the isotropic model of
Figure~\ref{fig:modelfitsone}.  When the best-fit models are then
passed through the Jeans equation to solve for the mass-to-light
ratio, we see the impact of anisotropy.  As shown in
Figure~\ref{fig:modelfitstwo}, the effects of varying the anisotropy
in this way dominate the statistical uncertainty from the range of
models that fit the data for a given anisotropy.  At our benchmark
radius of 5\Reff, the permitted range of mass-to-light ratios
is $\Upsilon_{{\rm 5},B}=6.8-12.2\Upsilon_{\odot,B}$, implying an
enclosed dark matter fraction of somewhere between none and 45\%,
which is still on the low side -- simple $\Lambda$CDM-based models
suggest that the value at these radii should be around 60\%
\citep{2005MNRAS.357..691N}.

Clearly, we could play the game of adopting ever more
complicated models for $\sigma_r(r)$ and $\beta(r)$ to try and fit
every observed feature, but the simplified nature of this entire
spherical Jeans-based analysis means that we are unlikely to learn
much more.  However, as one final embellishment, we consider a model
in which the orbits are isotropic at small radii, but undergo a
transition to a radial bias at large radii.  Such an arrangement was
the principal suggestion made by D+05 to explain the declining
velocity dispersion in NGC~3379, and, as discussed in
\S\ref{sec:dispsec}, is hinted at by the data themselves, as the
dispersion profile in Figure~\ref{fig:dispprof} seems to undergo a
transition between slopes at a radius of $\sim 70''$.
To model this arrangement, we adopt the functional form used by M{\L}05 of
\begin{equation}\label{eq:MLbeta}
\beta(r) = {0.5 r \over r+r_{\rm a}} , 
\end{equation} 
where  $r_{\rm a}\simeq 1.4$~\Reff\  $\simeq 65$''.
The Jeans analysis result from this anisotropy function is also
plotted in Figure~\ref{fig:modelfitstwo}.  Perhaps unsurprisingly,
this result lies somewhere between the isotropic case and the $\beta =0.5$ case.  
It implies a mass-to-light ratio within the benchmark
radius of $\Upsilon_{{\rm 5}, B} \simeq 10.6\Upsilon_{\odot,B}$,
corresponding to a dark matter fraction of $\simeq 30\%$.  Again,
there is evidence for dark matter, but not the dominant amount that 
$\Lambda$CDM models predict.  It also conflicts with the results from
the D+05 simulations that motivated it, which have a dark matter
fraction within this radius of $\simeq 60\%$.  We return to this
discrepancy in \S\ref{sec:squabble}.

\subsection{Comparison to Previous Data Modeling}
Before we look at the apparent conflict between mass models motivated
by simulations and those that result from the analysis of real data,
it makes sense to look a little more closely at the credibility of the
data modeling.  In particular, it is useful to explore how
reliably different analyses reproduce the same answers, and how
consistent the results obtained from different types of data are.

\begin{figure}
\includegraphics[width=8.6cm]{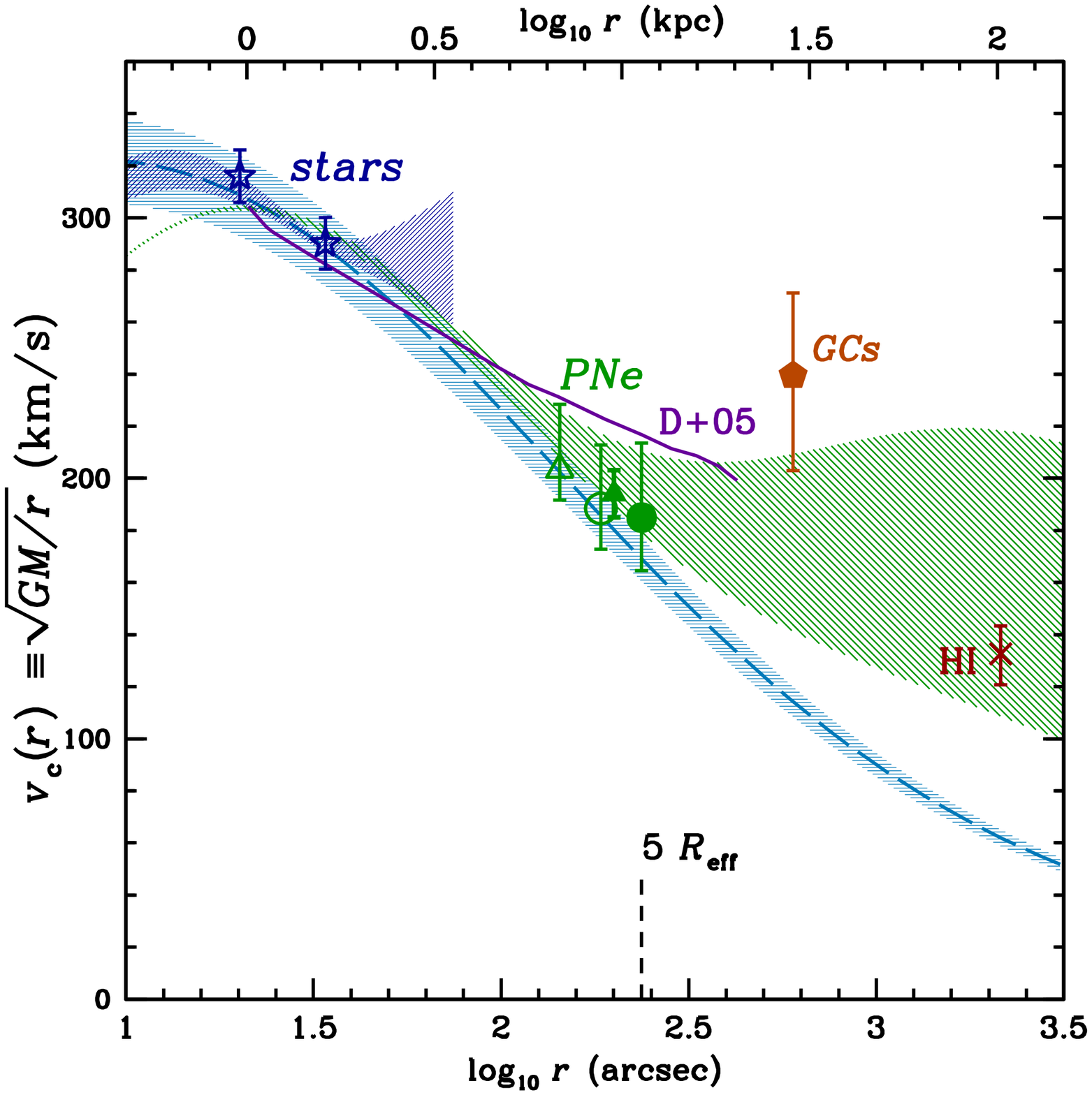}
\caption{ 
Summary of models determining the NGC~3379 circular velocity
profile.  From stellar absorption-line data, axisymmetric three
integral models from \citet{2006MNRAS.370..559S} and
\citet{2000AJ....119.1157G} are shown as stars, and the range of
spherical models determined by \citet{2000A&AS..144...53K}
based on distribution functions is shown as the inner shaded region.
Previous results from Jeans analyses of PNe are shown as a filled
triangle \citep{1993ApJ...414..454C}, an open circle (R+03), and an
open triangle \citep{2006AJ....131.2089S}. A combined orbit-model
fit to PNe and stellar data (R+03) is shown as the extended shaded
region.  A globular cluster result from \citet{2006A&A...448..155B}
and the \Hone{} circular speed from \citet{1985ApJ...288L..33S} are
also shown.  The result from this paper is shown as a filled circle.
The dashed line and surrounding shaded error region shows the
contribution to the circular velocity from the stellar component.
 The solid line shows the average result from the simulations of
\citet{2005Natur.437..707D}.}
\label{fig:gcshi}
\end{figure}
A summary of the available stellar-kinematic mass modeling for
NGC~3379 is provided in Figure~\ref{fig:gcshi}, where we have
transformed the enclosed mass into a circular speed for more direct
comparison with the rotation curves of spiral galaxies,
and as a quantity independent of the distance and luminosity model.
As the
caption describes, this figure shows the results from stellar
absorption-line data at small radii and those obtained from PNe at
larger radii, as well as the combined analysis of both types of data
by R+03.

We have also plotted the recent globular cluster enclosed mass
estimate from \citet{2006A&A...448..155B}.  Since the globular cluster
population is more extended than the stellar component, it offers a
probe of the enclosed mass at even larger radii.  Unfortunately,
however, NGC~3379 has a fairly meager population of globular clusters,
so the statistics are not yet good, making it impossible
to disentangle mass effects from orbital anisotropy (which
could be completely different from the anisotropy in the stellar
orbits).  Further, the more extended nature of their distribution and
the small numbers available mean that it may be difficult to
decontaminate the sample of interloping NGC~3384 clusters.  

One further mass probe at even greater distances is provided by the
kinematics of the rotating ring of \Hone{} that is seen to surround this
galaxy at a radius of $\sim 100\,{\rm kpc}$ \citep{1985ApJ...288L..33S}.
Again, NGC~3384 confuses the issue somewhat since this ring encompasses
it as well as NGC~3379.  However, if we adopt the simplest approach of
dividing the mass between the two galaxies in the same ratio as their
luminosities, attributing two thirds of the mass to NGC~3379, we finally
obtain the outermost point shown in Figure~\ref{fig:gcshi}.  What is most
striking about this plot is the consistency of the results.  When
different observations of the same tracer are employed, the results are
very similar.  Equally, when data sets are analyzed using  different
methods, be they Jeans analysis, distribution function fitting or
combining a library of orbits, or making different assumptions about the
symmetry of the galaxy, very comparable results are obtained.  When
completely different tracers are employed, the results still match up to
within the respective errors.  Ultimately, all these experiments agree
that although there is strong evidence for dark matter on the largest
scales, within the benchmark radius of 5\Reff\ the circular velocity
curve is well matched to that inferred from the stellar component alone
(also shown in Figure~\ref{fig:gcshi}),  so the amount of dark matter
there is at most modest.

\section{Scantily-Clad or Fully-Clothed?}\label{sec:squabble}
The relative consistency of the various models derived from kinematic
data returns us to the question of why the conclusion from these
studies differs so significantly from that reached by working in the
opposite direction of generating ``observations'' from simulations,
and comparing them to the data.  Specifically, D+05 claimed that they
were able to reproduce the earlier \PNS\ results on the velocity
dispersion profile by ``observing'' a simulated galaxy merger that
took place in the framework of standard $\Lambda$CDM dark matter
halos.  
This contrasts with the findings of R+03 and \citet{2005MNRAS.357..691N}
that the data were not consistent with $\Lambda$CDM halos.

D+05 concluded that three factors were responsible for the discrepancy:
\begin{enumerate}
\item Stars are thrown into radial halo orbits during the merger process.
\item The triaxial nature of the galaxies can depress $\sigma_{\rm p}$
       at some viewing angles.
\item The PNe trace young stars formed in the merger rather than the old stars
       usually assumed.
\end{enumerate}
We therefore consider these possibilities in turn.

The first factor seems plausible, since any stars thrown out from near
the centre of a galaxy in a merger process must end up on fairly
radial orbits.  Indeed, it seems to be a generic expectation for
galaxy halos
\citep{2005MNRAS.364..367D,2006MNRAS.365..747A,2006NewA...11..333H},
although there are some dissenting views on even this basic point
\citep{2005MNRAS.357..753G,2005AIPC..804..333A}.  However, it is
certainly not the explanation for the discrepancy between the
simulations and the data-fitting results, as this possibility has been
fully incorporated in the data fitting process of R+03.  
In fact, the anisotropy profile $\beta(r)$ found in this way was
very similar to that of the simulations.
So there is something else at work here besides the obvious effects
of radial anisotropy.

The second factor, a possible unfavorable viewing angle for NGC~3379, is
harder to rule out on the basis of observations of a single object.
Indeed, it is known that a flattened system viewed face on will have its
mass systematically underestimated if it is assumed to be spherical
\citep{2001MNRAS.322..702M}.  However, similar results were discussed by
R+03 in three other galaxies (NGC~821, 4494, and 4697). We would have to
be very unlucky for all these systems to be face-on, especially since
some of them appear significantly flattened on the sky.  

The final factor invoked by D+05 also does not seem particularly
plausible.  Although there is a prediction that the bright end of the 
PNLF is populated by younger stars  \citep{2004A&A...423..995M}, and one claim of the 
detection of such a system \citep{2006AJ....131..837S}, this
phenomenon has not been seen elsewhere.  This lack of
variation is particularly definitive in the best-studied case of our
neighboring spiral M31 where one might expect recent star formation
events to produce dramatic variations in the PN properties, yet no
differences at all are found in the bright end cut-off of the PN
luminosity function even when comparing the properties of bulge and
disk populations at the level of subtlety afforded by a sample of
almost 2800 PNe \citep{2006MNRAS.369..120M}.  The apparent
universality of the PN luminosity function in stellar populations of
all ages is completely inexplicable if the bright end of this function
is only populated where there are relatively young stars.

\begin{figure}
\includegraphics[width=8.6cm]{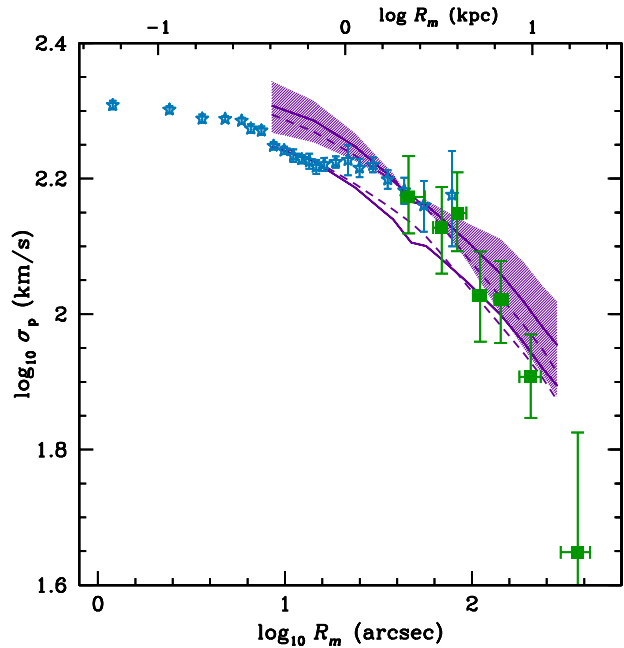}
\caption{ Comparisons of projected velocity dispersion profiles
between observations and \citet{2005Natur.437..707D} galaxy
simulations.  Symbols are as in Figure~\ref{fig:dispprof}, with the
addition of a solid line showing the dispersion profile of the old 
stars in the simulation, and a dashed line showing the profile for the young
stars.  Two possible normalizations are shown for each component.
The shaded area illustrates the region of scatter for one case.
}
\label{fig:dekelcomp1}
\end{figure}
 So, if none of these explanations holds water, why do the conclusions
derived from simulations differ so dramatically from those presented
here?  One issue is that the simulations do not reproduce
the observations as unambiguously as one
might want, and that there is significant uncertainty as to how the
simulation should be normalized to best match the data.
Figure~\ref{fig:dekelcomp1} compares the simulation's velocity
dispersion profile to that of the data.  It is apparent from this
figure that the over-all slope of the simulation's profile does not
match the data particularly well: the power-law index for the outer parts of the
simulation lies in the range 0.2 -- 0.4, as compared to the value of
0.4 -- 0.8 that we found for the data in \S\ref{sec:dispsec}.  The
match is a bit better to just the young stars in the simulation, but,
as discussed above, it is difficult to see why the detected PNe should
be associated with such a sub-population.  Even where there is a crude
match to the slope of the profile, the simulations do not reproduce
the structure seen in the data, so there is an ambiguity in the
velocity normalization depending on the radius at which one chooses to
match the two, as Figure~\ref{fig:dekelcomp1} illustrates.  This
uncertainty is compounded by the fact that there is a similar
ambiguity in the normalization of simulation and data in the spatial
coordinate as well: as was mentioned in \S\ref{sec:stellarmod}, the
observational measurement of NGC~3379's effective radius can depend
significantly on the manner in which it is determined.  Coupling this
uncertainty with the systematically different way in which \Reff\
 was determined from the simulations (based on the radius within
which half the mass is projected) means that matching the spatial
scale of the simulation to the data generates an additional
significant source of uncertainty.

The differences between the simulations and NGC~3379 extend beyond
the kinematics.  We note that the D+05
simulations contain a central blue ``hump'' in the stellar population
(D+05 Figure~2a) that is not seen in NGC~3379.  This phenomenon arises
from the well-documented problem of the over-collapse of baryons in
simulations of gas-rich mergers \citep[e.g.,][]{2003ApJ...590..619M}.
Although D+05 state that their results are unaffected by this feature,
it does raise a concern as to what subtler phenomena the simulation
might also be failing to match.  Since these systematically-incorrect
gas dynamics will very strongly affect the motions of any new stars that
form, it adds a  further question mark to any conclusions
based just on the young star population, although, as extensively
discussed, there are already many question marks hanging over the
association of the detected PNe with such a population.  
Additionally, this extra light adds an extra uncertainty in matching up the observed
photometry to that of the simulation, compounding the issue associated
with matching the spatial scales of the two.
And finally, dissipation in mergers is strongly linked to the
radius and velocity dispersion of the remnant's center
\citep{2006ApJ...650..791C},
so inaccuracies in this arena could very well lead to errors in
the slope of the dispersion profile.

\begin{figure}
\includegraphics[width=8.6cm]{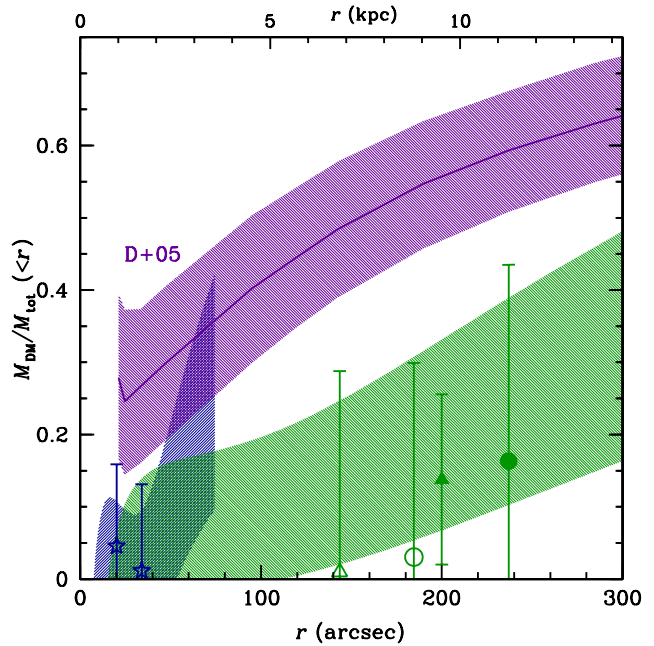}
\caption{The enclosed dark matter fraction as a function of radius for
the D+05 simulations, compared to the results derived directly from
observations of NGC~3379.  Symbols for the data are as in
Figure~\ref{fig:gcshi}.
}
\label{fig:dekelcomp2}
\end{figure}

Not withstanding these problems in matching the simulations to the NGC~3379 data,
we can still make a direct comparison between the mass profile results of
R+03 and of D+05 (Fig.~\ref{fig:gcshi}).  Although the relative normalization is
ambiguous, one can obtain a reasonably close match, given the run-to-run scatter
of the simulations.  So if the total mass profiles of R+03 and D+05 are not so
different, why do the conclusions about dark matter differ?
The answer lies in the decomposition of the total mass profile into
stellar and dark matter components.
As Figure~\ref{fig:dekelcomp2} shows, the dark matter fraction in the simulations
differs by more than a factor of two from that inferred from the data over a wide
range of radii, and requires that the galaxy have a sizeable 
 ($\sim 20$\%) dark matter fraction
at even the smallest radii.
This discrepancy can be largely traced to the different values adopted for
the $B$-band stellar mass-to-light ratio, $\Upsilon_{*,B}$.
In order to match up the inner rotation curve of the simulations to those derived
from the data, we find that the simulations only work if
$\Upsilon_{*,B} = 4 - 5\,\Upsilon_{\odot,B}$. This value clearly attributes less
mass to the stellar component (and hence more mass to the dark halo) than the
value in equation~\ref{eq:stellarML} adopted in this paper, which was
derived from stellar population synthesis results.  Although the lower
value is not completely implausible for this galaxy
\citep{2001AJ....121.1936G}, it is not really satisfactory to assign a
mass-to-light ratio in this ad hoc manner, and it is surely preferable
to adopt an astrophysically-motivated and constrained estimate.

The bottom line here is that a significant conflict exists between the
results derived from modelling the data directly and those obtained by
matching the data to simulations, and it is not clear which analysis is
flawed. The  Jeans modeling has the advantage that, by construction, the
data fit the model very well, but has the disadvantage that the model
itself, with its imposed spherical symmetry and simple view of the
tracer population, does not provide an entirely realistic picture of a
galaxy.  Conversely, the simulations are constructed to be as realistic
as possible in describing the properties of a galaxy, but then are not
designed to reproduce any particular galaxy with great accuracy. 
Ultimately, the solution must be to keep working from both directions,
producing more sophisticated dynamical models from the data and putting
better physics into a broader suite of simulations to match the
properties of particular systems.  Until these processes meet somewhere
in the middle, we really will not be able to say definitively whether
NGC~3379 is as lacking in dark matter as the modeling of this paper
suggests, or whether it has the ample dark-matter clothing that D+05
found in their simulations.

\section{Summary and Outlook}\label{sec:summary}

This paper presents the first fully-processed results from the \PNS\
survey of nearby ordinary elliptical galaxies, which has yielded a
catalog of 191 PNe in NGC~3379. The data processing is discussed in
detail. Given the novel nature of the instrument, we have gone into the
description  of the data processing in some detail.  Comparison with
existing smaller data sets has
  confirmed that the resulting kinematic measurements have velocity
  errors well within what is required for studying the large-scale
  dynamics of elliptical galaxies.
The data on NGC~3379 
have allowed us to study the kinematics of this
galaxy out to almost ten effective radii.
We detect a small amount of rotation, but random 
motions dominate throughout, the velocity
dispersion showing the regular pattern
expected for a near-spherical galaxy.
There is no evidence for
the large amounts of angular
momentum stored in rotation at large radii that are predicted by merger
simulations of elliptical formation.  

We  are able to confirm and extend the
previous findings that the velocity dispersion falls
with radius in a near-Keplerian manner. A spherical
Jeans model confirms that an isotropic model with a modest amount of
dark matter at large radii can fit the data very well.  These results
are found to be consistent with previous studies of the stellar and PN
kinematics of the galaxy, and also fit reasonably well with the mass
inferred from other tracers.  The amount of dark matter is low
compared to the predictions of $\Lambda$CDM, and, contrary to
previous claims, this discrepancy cannot be resolved simply by
changing the orbital anisotropy of the tracer population.
 
Our results continue to conflict with those found through comparison
with merger simulations by D+05, which claimed that the observed
kinematics are consistent with the presence of a conventional dark
halo around the galaxy.  
We have explored the various explanations
  that have been advanced to explain this discrepancy, but do not find
  any to be entirely convincing.  Just as there are shortcomings in the
  spherical Jeans modeling of the data, so there are issues in matching
  a simulation to a specific galaxy, and it is not clear which is more
  likely to be in error.  One interesting difference is that the D+05
  model contains significant amounts of dark matter even at very small
  radii, whereas the Jeans-based models do not.  We have traced much of
  this difference to the mass-to-light ratios adopted for the stellar
  component, which differ by almost a factor of two.  We thus appear to
  be arriving at the same ambiguity that beset attempts to decompose
  rotation curves of spiral galaxies into luminous and dark components,
  in that the amount of mass to attribute to the stellar component was
  not well constrained.  Although this issue remains a challenge, it is
  not necessarily insurmountable, and we hope to have made at least a
  first step toward addressing it by setting the mass-to-light ratio of
  the stellar component on the basis of astrophysically-motivated
  population synthesis models matched to the photometry of the
  individual galaxy, rather than leaving it as a variable parameter with
  which to re-enforce our prejudices.

In order to advance the data modelling approach to the study of
elliptical galaxies, to try to reconcile the results with those from
simulations, we propose to follow a two-pronged strategy.  First, we
intend to perform more sophisticated modeling of the data.  By
relaxing some of the constraints such as spherical symmetry, we will
be able to see what effect these artificial limitations might be
having on the inferences that we draw.  Further, we can make more
efficient use of the data than simply binning velocities to determine
a dispersion profile: with more detailed analysis, we can use the
complete projected phase space distribution, and can also extract the
information in a manner that is more robust than taking moments.
Second, we will carry out a comparable analysis on the other ordinary
elliptical galaxies for which we are obtaining data with \PNS.  By
analyzing a broader sample in a coherent and consistent manner, we
will be able to see just how common systems like NGC~3379 are, thus
ruling out some of the less interesting explanations for its strange
properties such as an unfortunate viewing angle.  We will also be able
to start to investigate how the large-scale dynamical properties of
elliptical galaxies correlate with other factors such as their
environment.  Hopefully, this further work will answer unequivocally
some of the fundamental questions that remain outstanding in the study
of the most ordinary of elliptical galaxies.

\acknowledgements

We would like to thank the Isaac Newton Group staff on La Palma for supporting
the \PNS\ over the years.
We also thank 
Michele Cappellari,
Walter Dehnen,
Avishai Dekel,
Wolfgang Gieren,
Gary Mamon, 
and
Arend Sluis
for helpful comments, conversations, and contributions.

AJR is supported by the FONDAP Center for Astrophysics
CONICYT 15010003.  MRM is supported by a PPARC Senior Fellowship,
LC by a grant from the NWO of the Netherlands, and
NRN is funded by CORDIS within FP7 with a Marie Curie European
Reintegrational Grant, contr. n. MERG-6-CT-2005-014774.

\bibliography{ndouglas}

\end{document}